\documentclass[fleqn,usenatbib]{mnras}

\usepackage{newtxtext,newtxmath}

\usepackage[T1]{fontenc}
\usepackage{ae,aecompl}
\usepackage{siunitx}
\usepackage{xspace}

\DeclareRobustCommand{\VAN}[3]{#2}
\let\VANthebibliography\thebibliography
\def\thebibliography{\DeclareRobustCommand{\VAN}[3]{##3}\VANthebibliography}

\usepackage{graphicx}	
\usepackage{amsmath}	
\usepackage{xspace}

\newcommand{\BGSB}{\uppercase{bgs bright}\xspace}
\newcommand{\BGSF}{\uppercase{bgs faint}\xspace}
\newcommand{\BGS}{\uppercase{bgs}\xspace}
\newcommand{\BASS}{\uppercase{bass}\xspace}

\newcommand{\BS}{\uppercase{bs}\xspace}

\newcommand{\DECaLS}{\uppercase{dec}a\uppercase{ls}\xspace}

\newcommand{\FRACMASKED}{{\uppercase{fracmasked}}\xspace}
\newcommand{\FRACFLUX}{{\uppercase{fracflux}}\xspace}
\newcommand{\FRACIN}{{\uppercase{fracin}}\xspace}
\newcommand{\GAMA}{\uppercase{gama}\xspace}
\newcommand{\GAIA}{{\it Gaia}\xspace}

\newcommand{\LS}{\uppercase{ls}\xspace}
\newcommand{\LG}{\uppercase{lg}\xspace}
\newcommand{\LSLGA}{\uppercase{lslga}\xspace}

\newcommand{\MzLS}{\uppercase{m}z\uppercase{ls}\xspace}

\newcommand{\NGC}{\uppercase{ngc}\xspace}

\newcommand{\NQ}{\uppercase{nq}\xspace}
\newcommand{\PSF}{\uppercase{psf}\xspace}

\newcommand{\QCs}{\uppercase{qc}s\xspace}

\newcommand{\SDSS}{\uppercase{sdss}\xspace}
\newcommand{\SGC}{\uppercase{sgc}\xspace}

\newcommand{\TRACTOR}{\uppercase{tractor}\xspace}

\newcommand{\REX}{\uppercase{rex}\xspace}

\newcommand{\degsq}{deg$^{2}$}

\RequirePackage[normalem]{ulem} 
\RequirePackage{color}\definecolor{RED}{rgb}{1,0,0}\definecolor{BLUE}{rgb}{0,0,1} 

\title[DESI BGS using DR9 Legacy Imaging Surveys]{Preliminary clustering properties of the DESI BGS bright targets using DR9 Legacy Imaging Surveys}

\author[P. Zarrouk et al.]{
Pauline Zarrouk$^{1,2}$\thanks{E-mail: pauline.zarrouk@lpnhe.in2p3.fr},
Omar Ruiz-Macias$^{2,3}$,
Shaun Cole$^{2}$,
Peder Norberg$^{2,4}$,
Carlton Baugh$^{2,3}$
\newauthor
David Brooks$^{5}$,
Enrique Gazta\~naga$^{6}$, 
Ellie Kitanidis$^{7}$,
Robert Kehoe$^{8}$,
Martin Landriau$^{9}$,
\newauthor
John Moustakas$^{10}$,
Francisco Prada$^{11}$,
Gregory Tarl\'e$^{12}$
\\
\scriptsize $^{1}$ Sorbonne Universit\'e, Universit\'e Paris Diderot, CNRS/IN2P3, Laboratoire de Physique Nucl\'eaire et de Hautes Energies, LPNHE, 4 place Jussieu, F-75252 Paris, France \\
\scriptsize $^{2}$ Institute for Computational Cosmology, Department of Physics, Durham University, South Road, Durham DH1 3LE, UK \\
\scriptsize $^{3}$ Institute for Data Science, Durham University, South Road, Durham DH1 3LE, UK \\
\scriptsize $^{4}$ Centre for Extragalactic Astronomy, Department of Physics, Durham University, South Road, Durham DH1 3LE, UK \\
\scriptsize $^{5}$ Department of Physics \& Astronomy, University College London, Gower Street, London, WC1E 6BT, UK \\
\scriptsize $^{6}$ Institute of Space Sciences (ICE, CSIC), Campus UAB, Carrer de Can Magrans, s\/n, 08193 Bellaterra (Barcelona), Spain \\
\scriptsize $^{7}$ Department of Physics, University of California, Berkeley, 366 LeConte Hall, Berkeley, CA 94720, USA \\
\scriptsize $^{8}$ Department of Physics, Southern Methodist University, 3215 Daniel Ave., Dallas, TX, 75205, USA \\
\scriptsize $^{9}$ Lawrence Berkeley National Laboratory, 1 Cyclotron Road, Berkeley, CA 94720, USA \\
\scriptsize $^{10}$ Department of Physics and Astronomy, Siena College, 515 Loudon Road, Loudonville, NY 12211, USA \\
\scriptsize $^{11}$ Instituto de Astrofısica de Andalucıa (CSIC), Glorieta de la Astronomıa, s/n, E-18008 Granada, Spain \\
\scriptsize $^{12}$ Department of Physics, University of Michigan, Ann Arbor, MI 48109, USA
}

\date{Accepted XXX. Received YYY; in original form ZZZ}

\pubyear{2021}

\begin{document}
\label{firstpage}
\pagerange{\pageref{firstpage}--\pageref{lastpage}}
\maketitle

\begin{abstract} We characterise the selection cuts and clustering properties of a magnitude-limited sample of bright galaxies that is part of the Bright Galaxy Survey (BGS) of the Dark Energy Spectroscopic Instrument (DESI) using the ninth data release of the Legacy Imaging Surveys (DR9).  We describe changes in the DR9 selection compared to the DR8 one as explored in Ruiz-Macias et al. (2021). We also compare the DR9 selection in three distinct regions: BASS/MzLS in the north Galactic Cap (NGC), DECaLS in the NGC, and DECaLS in the south Galactic Cap (SGC). We investigate the systematics associated with the selection and assess its completeness by matching the BGS targets with the Galaxy and Mass Assembly (\GAMA) survey.
We measure the angular clustering for the overall bright sample ($r_{\rm mag} \leq 19.5$) and as function of apparent magnitude and colour. This enables to determine the clustering strength $r_{0}$ and slope $\gamma$ by fitting a power-law model that can be used to generate accurate mock catalogues for this tracer. We use a counts-in-cells technique to explore higher-order statistics and cross-correlations with external spectroscopic data sets in order to check the evolution of the clustering with redshift and the redshift distribution of the BGS targets using clustering-redshifts. While this work validates the properties of the BGS bright targets, the final target selection pipeline and clustering properties of the entire DESI BGS will be fully characterised and validated with the spectroscopic data of Survey Validation.
\end{abstract}




\section{Introduction}

Building ever larger and more  accurate 3D maps of the large-scale structure of the Universe is the key to understanding the formation and evolution of the Universe, and, in particular, the late-time acceleration of its expansion. However, the mechanism which is responsible for this acceleration remains a major unknown in modern cosmology. In order to account for this discovery, the cosmological constant $\Lambda$ was re-introduced in Einstein's equations of General Relativity, such that it represents the major component of the current energy content of the Universe. The observations could also be explained by more complex dark energy models with time-dependent properties or even a modification of General Relativity on cosmological scales, (e.g. \citealt{LinderCahn07,Guzzo+08, Alam:2020}). 
 
The Dark Energy Spectroscopic Instrument (DESI, \citealt{DESI2016:surveys}) has been designed to meet the goal of elucidating the cosmic acceleration and it is the first of the new generation of galaxy surveys, the  so-called Stage IV dark energy experiments using the terminology of  the Dark Energy Task Force report~\citep{Albrecht:2006}, that has already started taking data. DESI is a multi-fiber spectrograph installed on the Mayall 4-meter telescope at Kitt Peak, Arizona, in the United States, which can collect 5,000 spectra simultaneously using robot fiber positioners. 

In order to create these 3D maps, we first require imaging surveys that provide the targets we are going to put a fiber on with the DESI instrument so that we can extract accurate and reliable positional information in the radial direction. We use three different imaging surveys that provide optical data over 20,000 deg$^2$ for DESI targeting and constitute the Legacy Imaging Surveys~\citep{Dey+19}: i) the Beijing-Arizona Sky Survey (BASS) imaged regions at $\it {Dec} \geq +32$ deg in the Northern Galactic Cap (NGC), in the $g$ and $r$ bands, ii) the Mayall z -band Legacy Survey (MzLS) imaged the $\it {Dec} \geq 32$ deg region and iii) the DECaLS program that made use of other DECam data within the Legacy Surveys (LS) footprint. 

DESI will conduct  dark time and bright time programs in parallel. During the dark time program four main target classes will be observed: about 6 million luminous red galaxies (LRG) in the redshift interval $0.4 < z < 1.0$, 17 million [OII] emission-line galaxies (ELG) in $1.0 < z < 1.6$ and 2.5 million quasars (QS0), where quasars with $z < 2.1$ will serve as direct tracers of the matter density field and quasars with $2.1 < z < 3.5$ will provide Lyman-$\alpha$ absorption features in order to probe the distribution of neural hydrogen in the intergalactic medium. During bright time, when lunar conditions are too bright for good observations of these faint objects, DESI will obtain 10 million spectra of bright galaxies up to $z \sim 0.4$ (Bright Galaxy Survey, BGS) and a sample of local stars (Milky Way Survey, MWS). The DESI footprint will cover 14, 000 deg$^2$.

Here, we focus on the DESI BGS program that will likely be comprised primarily of galaxies selected to a limiting magnitude. The dark energy science goals of the BGS are to produce a data set that allows the best achievable measurements of Baryon Acoustic Oscillations, (BAO, \citealt{Eisenstein+05,Cole+05}) and Redshift-Space-Distortions  (RSD, \citealt{Jackson72,Kaiser87}) at redshifts $z \leq 0.4$. Although the statistical precision of BGS clustering measurements is usually lower than that of the dark-time surveys, working near $z = 0$ gives maximum leverage against higher redshift measurements and CMB constraints. The exceptionally high sampling density, low-redshift sample and the wide range of galaxy bias represented in the BGS offer new opportunities for innovative analysis techniques, such as multi-tracer methods that measure common Fourier modes of the density field using tracers of different bias  and tests of systematic effects as different tracers with different clustering properties should yield compatible cosmological constraints~\citep{McDonaldSeljak09,Blake+13,Ross+14}. The DESI BGS will also enable novel investigations of structure growth using galaxy groups and clusters or combination with galaxy-galaxy weak lensing. By spanning a wide range of galaxy properties, the DESI BGS will also be an extraordinary resource for studying the properties of galaxies and galaxy groups and the relations between galaxies and dark matter, advancing our understanding of how primordial fluctuations grow into the rich structure of galaxies observed today.

In order to achieve these science goals, the target selection pipeline, the imaging systematics and the resulting clustering properties of the BGS targets need to be tested and validated. The purpose of this paper is to present the implications of improvements in the imaging data processing in the context of selecting a magnitude-limited galaxy sample for spectroscopic observations of large-scale structure.  We present the motivation for several changes to selection criteria since the most recent study based on previous imaging reduction~\citep{Ruiz+20,Ruiz+21} and the clustering characteristics of the resulting magnitude-limited sample. The studies presented here rely only on the imaging data collected for identification of DESI spectroscopic targets;  final target selection for the BGS program will be presented in a future paper that presents the full interpretation of sample selection derived from early spectroscopic observations from DESI. A complementary study by \cite{Kitanidis+19} examined the impact of imaging systematics on the selection and clustering of targets in the LRG, ELG and QSO DESI surveys, using an earlier release of the Legacy Survey imaging data.

The paper is organised as follows. In Section~\ref{sec:data}, we present the imaging data and target selection cuts we use to select our targets that relies on the work we did first with DECaLS DR8 in our companion paper~\cite{Ruiz+21} and we investigate the potential systematics. In Section~\ref{sec:angular}, we present several angular clustering properties of the BGS targets and Section~\ref{sec:cross} is dedicated to the the cross-correlation measurements with external spectroscopic datasets. Our conclusion can be found in Section~\ref{sec:concl}.

\section{BGS target catalogue using DR9}
\label{sec:data}

A candidate BGS selection using DR8 DECaLS was first explored in~\citep{Ruiz+21} and also in a research note~\citep{Ruiz+20}. The core of this selection for the bright sample remains unchanged: a bright sample at $r < 19.5$ (BGS Bright) and a faint sample at $19.5 < r < 20$ (BGS Faint). From here out, in Section~\ref{sec:comp_dr8} we first describe the changes in the imaging pipeline for DR9 and the changes in the target selection cuts they lead and then we present a comparison between BASS/MzLS and DECaLS in Section~\ref{sec:bmzls_decals}. We quantify the impacts of the changes between DR8 and DR9 for the masking around bright stars in Section~\ref{sec:xstars} and for the large galaxies in Section~\ref{sec:xLG}.


\subsection{Changes between DR8 and DR9}
\label{sec:comp_dr8}

\subsubsection{Changes in the imaging pipeline}

Compared to its predecessor, DR9 incorporates some major and minor changes that affect the photometry of the objects, and hence the  target selection. These changes are incorporated in 
legacypipe~\footnote{\url{https://github.com/legacysurvey/legacypipe}}, which is the DESI Imaging Legacy Surveys data reduction pipeline.
Below we list the most important changes relevant for \BGS~ target selection:

\begin{enumerate}
\item {\it Iterative detection}: After the first round of fitting, TRACTOR conducts a second round of detections over the data-model residuals with the aim of finding additional sources. 
\item {\it Extended PSF model}: An extended PSF model is used to subtract the wings of bright stars from DECam images only.
\item {\it Sersic fitting model}: The composite (COMP) morphological model has been replaced by a Sersic profile (SER). A source is classified
as SER if a Sersic profile
provides a better fit than other profiles, PSF, EXP and DEV, as quantified by a $\chi^{2}$ that takes account of additional free parameters of the Sersic fit.
\item {\it Relaxed \GAIA \, PSF criterion}: TRACTOR forces \GAIA~objects to be fitted as PSF GAIA sources if they meet the condition ($G \leq 18$ $\&$ AEN $< 10^{0.5}$) OR ($G \leq 13$), that was previously set to ($G \leq 19$ $\&$ AEN $< 10^{0.5}$) OR ($G \leq 19$ $\&$ AEN $< 10^{0.5+0.2(G-19)}$). Here $G$ is GAIA DR2 $G$-band magnitude, and AEN is the \GAIA \, astrometric excess noise parameter.
\item {\it Pre-fitting for large sources}: Regions around large galaxies and globular clusters have their own local source extraction, which is performed separately from the normal TRACTOR run. The parent catalogues of these objects have improved extensively since DR8 and this constitutes the Siena Galaxy Atlas 2020~\footnote{\url{https://sga.legacysurvey.org}} (SGA 2020 \citealt{SGA2020}).
\end{enumerate}

\subsubsection{Changes in the BGS target selection}
\label{sec:changeTS_dr8_dr9}

These improvements led to changes in the target selection pipeline that are summarised below: 
\begin{enumerate}
    \item {\it Bright stars}: the size of the masking radius around bright stars is reduced by a factor of 2. In Section~\ref{sec:xstars}, we investigate the consequences of such a change in the BGS selection
    \item {\it Large galaxies}: the size of the masking radius around large galaxies in DR9 has increased from roughly 10 arsec to 30 arcsec, however the pre-fitting for large sources led to major improvements in the photometry extraction of these objects which allows us not to apply this mask in DR9. Moreover, we remind the reader that in~\citep{Ruiz+21}, we assess the completeness of the previous selection using DR8 by matching the BGS targets with GAMA DR4 and found that we were missing about 25 true galaxies/\degsq.
    \item {\it Quality cuts (QC)}: the "quality cuts" (QC) defined in~\citep{Ruiz+21} for \DECaLS~DR8 are cuts in $\texttt{FRACMASKED}\_i < 0.4$, $\texttt{FRACIN}\_i  > 0.3$, $\texttt{FRACFLUX}\_i  < 5$, where $i \equiv\textrm{$g$, $r$ or $z$}$. \texttt{FRACIN} is used to select sources for which a large fraction of the model flux lies within the contiguous pixels to which the model was fitted, \texttt{FRACFLUX} is used to reject objects that are swamped by flux from adjacent sources, and \texttt{FRACMASKED} is used to veto objects with a high fraction of masked pixels. The overall improvement in the quality of the photometry assessed in next sections~\ref{sec:check} and~\ref{sec:vi} allows us to adopt a less conservative definition of the QC. Instead of applying these cuts in the three bands, we require only two out of three bands (see Eq.~\ref{eq:qualcut}). For posterior analysis, we will refer to DR8 QC as old FRACS, and for the QC in this work as just new FRACS. The new FRACS is a subset of around $1/3$ of the old FRACS. In Section~\ref{sec:check}, we assess again the completeness with respect to GAMA and we remind the reader that for DR8 we were missing about $70$ true galaxies/\degsq~because of old FRACS. In order to assess the current selection given by equation~\ref{eq:qualcut}, we perform an imaging visual inspection (VI) with examples of targets flagged by the old FRACS. The details of the VI set up and results are given in Section~\ref{sec:vi}.

\end{enumerate}

\begin{eqnarray}
\textrm{FRACMASKED}\_i &&< 0.4, \nonumber \\
\textrm{FRACIN}\_i  &&> 0.3, \nonumber \\
\textrm{FRACFLUX}\_i  &&< 5, \nonumber \\
\textrm{where} \ i &=& \textrm{$\{g,r\}$, $\{g,z\}$ or $\{r,z\}$}.
\label{eq:qualcut}
\end{eqnarray}

Compared with the \BGS~selection defined for \DECaLS~DR8, the current \BGS~selection with DR9 increases the overall target density by $\sim20$ objects/\degsq. Table~\ref{tab:bgs_density} shows the gain in target density for \BGSB~and \BGSF~and for the three regions, together with the resulting target density for this selection and the corresponding effective area after accounting for the spatial masking.

\begin{table}
\caption{Target density in objects/\degsq ($\eta$), increase in target density in objects/\degsq ($\eta$) of the current \BGS~ target selection (DR9) compared with the \BGS~ selection defined for \DECaLS~ DR8 and the effective area ($A_\textrm{eff}$) in \degsq~ after accounting for the spatial masking. We show results for \BGSB~ and \BGSF~ each divided into  three regions; \BASS/\MzLS, \DECaLS~ \NGC~ and \DECaLS~ \SGC.}
\label{tab:bgs_density}
\centering
\begin{tabular}{ |p{1.5cm}||p{1.5cm}|p{1.8cm}|p{1.8cm}|  }
 \hline
 BGS &  BASS/MzLS &  DECaLS NGC &  DECaLS SGC \\
 \hline
 $\eta_\textrm{bright}$ &      813.4 &       875.6 &       839.0 \\
 $\eta_\textrm{faint}$ &      569.4 &       598.8 &       581.1 \\
 \hline
 $\Delta\eta_\textrm{bright}$ &      +12.7 &        +14.1 &        +12.5 \\
 $\Delta\eta_\textrm{faint}$ &      +\hphantom{0}7.7 &         +\hphantom{0}8.2 &         +\hphantom{0}7.3 \\
 \hline
 $A_\textrm{eff}$ &      4493 &       5263 &       4326 \\
 \hline
\end{tabular}
\end{table}



\subsubsection{Visual inspection on the imaging}
\label{sec:vi}
The goal of perform an imaging VI is to assess our selection and reduce human uncertainties in the classification by involving as many people as possible. For this purpose, we create the LSVI\footnote{\url{https://lsvi-webtool.herokuapp.com/}} (Legacy Surveys Visual Inspection), an interactive web framework that automatize this process. LSVI essentially create postage stamps galleries of the Legacy Surveys Sky Viewer web site viewer\footnote{Legacy Surveys / D. Lang (Perimeter Institute) \url{legacysurvey.org/viewer}}~(D. Lang in prep.) with their corresponding classification labels.

To assess the the conservative selection of the QC we adopt for DR9, we took a sparsely selected sample of $2000$ \BGS~ targets in a $\sim420$ \degsq area in \DECaLS~ DR9 flagged by the old FRACS. This sample accounts for $\sim 35$ per cent out of the total available in this area. The classification labels are, GAL: The object is a galaxy; STAR The object is a star; BT: The object is contaminated by (or is a fake source from) a bleed trail; DS/H: The object is contaminated by (or is a fake source from) a diffraction spike or a star halo; FRAG: The object is spurious from a fragmented large galaxy; JUNK: The object does not fit in any of the previous classifications but it is clearly spurious; UNK: The object does not fit in any of the previous classifications and it's unclear whether it is a galaxy or not; NI: The object has not yet been inspected.

A total of six people participated in the classification, and the results where classified in three categories: i) confirmed galaxy (CG): two or more people say the object is a galaxy and less than two people say it is anything but a galaxy; ii) confirmed non-galaxy (CNG): two or more people say the object is anything but a galaxy and less than two people say it is a galaxy; and iii) inconsistent classification (IC): is neither of previous categories. Out of the $2000$ objects, $35$ per cent are CG, $43$ per cent are CNG, and $22$ per cent are IC. The CNG include stars and other artifacts such as bleed trails, diffraction spikes, stellar halos and fragmented images. There is no easy way to isolate the CG while getting rid of spurious sources. However, for the subset of these old FRACS, the new FRACS defined by equation~\ref{eq:qualcut}, results show that this subset get rid of only around $10$ per cent of the CG, around $60$ per cent of the CNG, and around $30$ per cent of the IC.

\subsubsection{Validation with GAMA}
\label{sec:check}

In DR8 DECaLS, we assess the completeness of our \BGS~catalogue with respect to \GAMA, whose main target sample~\citep{10.1093/mnras/stx3042} is highly complete for galaxies with $14 < r_\textrm{SDSS} < 19.8$, where $r_\textrm{SDSS}$ is the \SDSS~Petrosian $r$-band magnitude, with $98.85$\% of the objects in the catalogue having good redshifts with a quality flag \NQ$\geq 3$, after applying a  redshift cut at $z > 0.002$ to remove the remaining stars. We match the \BGS~targets with the \GAMA~Main Survey DR4 galaxy catalogue\footnote{This is an unreleased version that the \GAMA~collaboration made available to us. It is essentially the same as \GAMA~DR3, but with more redshifts.} using a maximum linking length of $1$ arcsec and we focus on three of the five \GAMA~fields: G09, G12, G15.
First, we cross-matched the \GAMA~catalogue with the Legacy Imaging Surveys DR9. This results in a matched catalogue of $1007.5$ objects/\degsq \, in a $178$ \degsq~area,  which represents $99.8$ per cent of all the \GAMA~galaxies. We define four different \BGS~selections to test the completeness with respect to \GAMA. The four samples are: i) DR8 cuts (nominal DR8), ii) nominal DR9 with no LG mask and new FRACS (current selection), iii) no LG mask and old FRACS, iv) no LG mask and no FRACS. From the total of $1007.5$ objects/\degsq \, matches between \DECaLS~and the \LS~DR9, the nominal DR9 and the no LG,  no FRACS samples yield the most \BGS~ targets with $1,003$ and $1,004$ objects/\degsq \, respectively. Next in order of number of matches comes the no LG, old FRACS sample with a matched \BGS~sample of $998$ objects/\degsq. Finally, the nominal DR8 sample gives the fewest matches with $995$ objects/\degsq.

\begin{figure}
	\centering
	\includegraphics[width=\columnwidth]{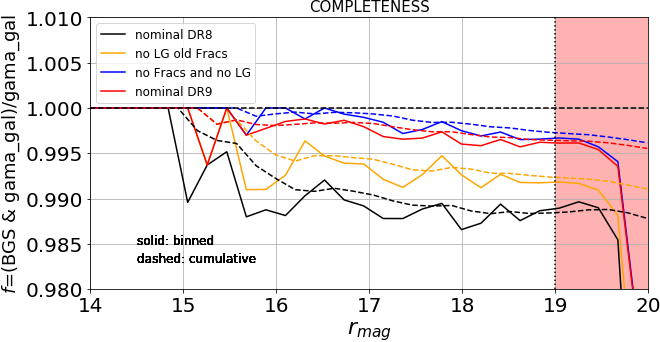}
    \caption{Completeness of the \BGS~targets according to different choices of target selection cuts with respect to \GAMA~which can be considered as complete for $14 < r_\textrm{\SDSS} < 19.8$.} 
    \label{fig:completeness} 
\end{figure}

Fig.~\ref{fig:completeness} shows the completeness with respect to \GAMA~of the four samples obtained by computing the ratio of \BGS~matched with \GAMA~over the \GAMA~objects. These results show that the sample with no LG and no FRACS has the highest completeness with respect to \GAMA, followed by the nominal DR9 sample; both these samples have more than $99.5$ per cent completeness. However, if we had taken the ratio of the \BGS~matched with \GAMA~over the total of \BGS~targets, we would find that the nominal DR9 sample has the highest completeness, which can be interpreted as being the sample with the least contamination. Focusing on our current \BGS~selection, the nominal DR9 sample, there are $4.5$ objects/\degsq~ from \GAMA~ that are missing in \BGS, of which $3.8$ objects/\degsq~are due to the star-galaxy separation, and the remaining $0.7$ objects/\degsq~comes from the \QCs. 
Based on these results, we can conclude that our current \BGS~sample has a completeness with respect to \GAMA~which is above 99.5\%.

\subsection{Comparison of DR9 DECaLS and DR9 BASS/MzLS}
\label{sec:bmzls_decals}

Looking at the survey depth in each band, characterised by the $5\sigma$ AB magnitude detection limit for a $0.45$ arcsec \REX~galaxy profile, the \DECaLS~$g$ and $r$ bands go deeper than the equivalent bands in \BASS. In the $z$- band, however, \DECaLS~and \MzLS~have a similar depth. Whilst, for the purpose of this work these depths are sufficient for the \BGS~selection, we are interested here to see how the magnitudes measured in the same bands differ between the surveys, which were conducted with different instruments at different telescopes.

\DECaLS~and \BASS/\MzLS~overlap in the \NGC~ at around $\it {Dec} = 32$ deg within $29 < \it {Dec} < 35\, \textrm{deg}$. For this analysis we looked at a $76$ deg$^{2}$ area in the region 200 deg < RA < 240 deg and 29 deg $< \it {Dec} <$ 35 deg. The area was computed using a random catalogue with density of $15,000$ objects per \degsq \, and a HEALPIX grid of side $1024$. In order to compare the photometry in \DECaLS~ and \BASS/\MzLS, we find all target matches within a distance of $0.5$ arcsec. To avoid incomplete regions, we require NOBS$_{i} > 0$ for $i = g, r$ and $z$ for the three surveys. After the match, we define the \BGS~ objects for the three surveys using and we find agreement between  $1,328$ BGS  objects/\degsq, and disagreement for $66$ objects/\degsq \, that are in  \BGS~in \DECaLS~ but which are not \BGS~ in \BASS/\MzLS; conversely there are $28$ BGS objects/\degsq~in \BASS/\MzLS~that are not in the \DECaLS~BGS.
    
\begin{figure*}
 \centering
 \includegraphics[scale=0.4]{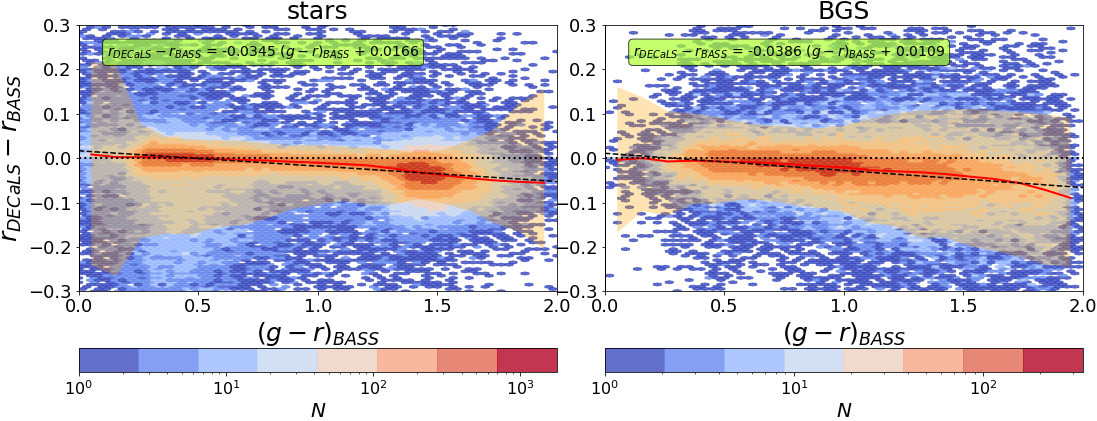}
 \caption{\DECaLS~ and \BASS/\MzLS~ matched Stars (left) and \BGS~ targets (right) showing $r_\textrm{DECaLS} - r_\textrm{BASS-MzLS}$ as a function of $(g-r)_\textrm{BASS}$ colour. Matched stars are shown in the left-hand side while \BGS~ in the right-hand side. Stars are as defined by our \GAIA star-galaxy classification. The colour bar indicates the number of objects within the hexagonal cell. The red solid line shows the median and the orange shaded region shows the the $3$ and $97$ percentiles. The black dashed lines show the best straight line fits to the median relation as given in the green boxes above.}
 \label{fig:photometry_decals_bass}
 \end{figure*}
 
 \begin{figure}
	\centering
	\includegraphics[width=\columnwidth]{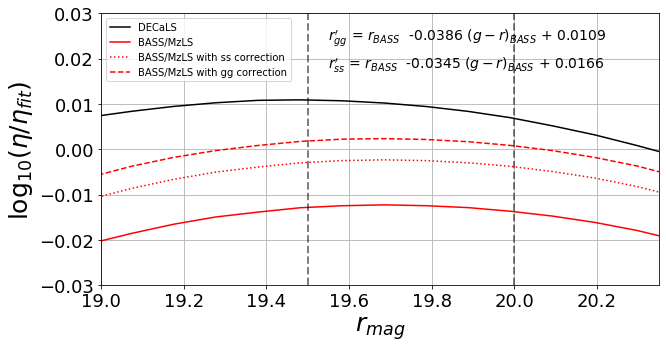}
    \caption{The \BGS target density ($\eta$) divided by the the target density of a fiducial linear fit, $\log_{10}(\eta_\textrm{fit}) = 0.46 \times r_\textrm{mag} + 6.10$. The solid black and red lines show the \DECaLS and  \BASS/\MzLS \BGS target densities within their full footprints. The red dashed and dotted lines shows  \BASS/\MzLS \BGS target density after applying the $r$-mag transformation equations~\ref{eq:transformations_gg} and ~\ref{eq:transformations_ss} respectively.}
    \label{fig:counts} 
\end{figure}

 We find that most of the disagreements are due to a shift in the $r$-band magnitude. Fig.~\ref{fig:photometry_decals_bass} shows $r_\textrm{DECaLS} - r_\textrm{BASS-MzLS}$ as a function of $(g-r)_\textrm{BASS}$ for two samples: one with only stars in both \BASS/\MzLS~and in \DECaLS, and the other with only \BGS~objects. We can fit the magnitude difference as a linear function in $(g-r)_\textrm{BASS}$. Equations~\ref{eq:transformations_gg} and~\ref{eq:transformations_ss} show these $r$-band photometry transformations of the  \BASS/\MzLS~system to the  \DECaLS~system for \BGS~ matches ($r^{\prime}_\textrm{gg}$) and for star matches ($r^{\prime}_\textrm{ss}$),  respectively
 
\begin{eqnarray}
r^{\prime}_\textrm{gg} &=& r_\textrm{BASS}  -0.039 (g-r)_\textrm{BASS} + 0.011 \label{eq:transformations_gg} \\
r^{\prime}_\textrm{ss} &=& r_\textrm{BASS}  -0.035 (g-r)_\textrm{BASS} + 0.017. \label{eq:transformations_ss} 
\end{eqnarray}

From equations~\ref{eq:transformations_gg} and \ref{eq:transformations_ss} we can see that the $r$-band magnitude in \BASS/\MzLS~is fainter than in  \DECaLS. The $r$-band magnitude offset is $r_\textrm{\BASS}-r_\textrm{\DECaLS} \approx 0.026$ for \BGS~ objects. In Fig.~\ref{fig:counts} we show the number counts as a function of $r$-mag for \BGS~ objects in both regions in the overlapping $76$~\degsq region and compare this with number counts of \BASS/\MzLS~after correcting the $r$-mag by applying  equations~\ref{eq:transformations_gg} and ~\ref{eq:transformations_ss}. These results show that both regions can achieve similar target densities if we apply a linear transformation in \BASS/\MzLS, increasing the overall target density to $1430$ objects/\degsq. Compared to the $1383$ objects/\degsq~without the colour correction, this represents a increase of $3.4$ per cent. 


\subsection{Bright stars}
\label{sec:xstars}

\begin{figure*}
	\centering
	\includegraphics[scale=0.4]{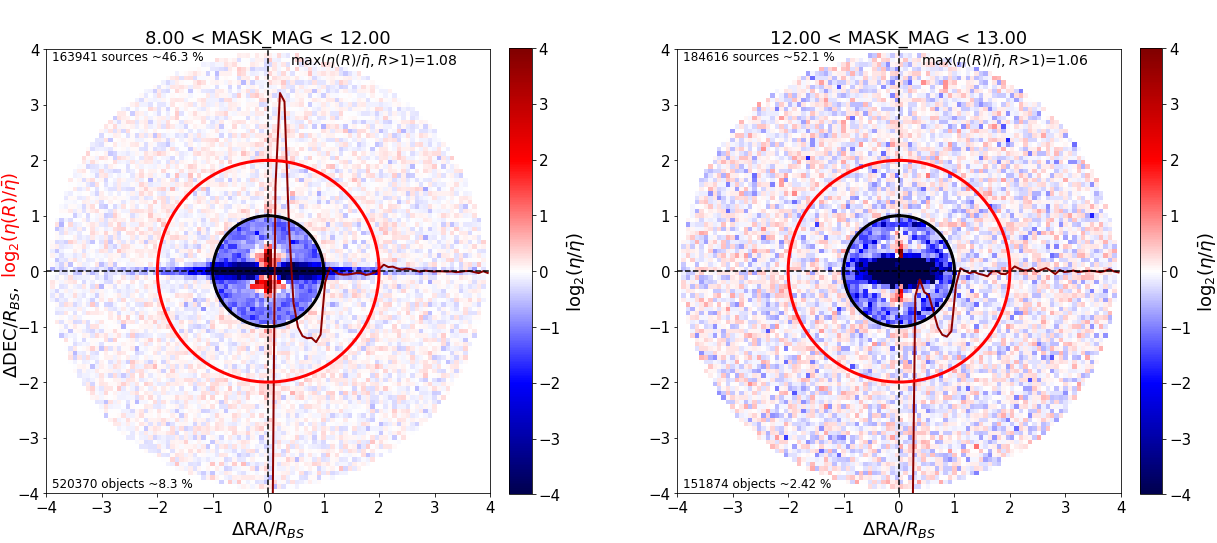}
    \vskip -0.2cm
    \caption{2D histograms of the positions of \BGS \, objects from BASS/MzLS relative to their nearest Bright Star (\BS) taken from the \GAIA \, and Tycho catalogues down to $G$-mag and visual magnitude \textsc{mag}\_\textsc{vt} of $13$ respectively. 
    These stacks are performed in magnitude bins in the bright stars catalogue from magnitude $8$ to $12$ (left) and $12$ to $13$ (right). The stacks are made using angular separations rescaled to the masking radius function of DR9 (inner black circle), which means that objects within a scaled radius of $0$ to $1$ will be masked out by the \BS veto while objects with $R = r/R_\textrm{BS} > 1$ will not (here $r^2 = (\Delta \textrm{RA}^2 \cos(\it {Dec})^2 + \Delta \it {Dec}^2$).
    The colour scale shows the ratio of the density per pixel ($\eta$) to the mean density ($\bar{\eta}$) within the shell $1.1 < r/R_\textrm{BS} < 4$. The density ratio is shown on a $\log_2$ scale where red shows over-densities, blue corresponds to under-densities and white shows the mean density. The outer red circle shows the masking radius of the LS DR8 data. The red solid line shows the radial density profile on the same scale as the colour distribution $\log_2(\eta(R)/\bar{\eta})$ where $\eta(R)$ is the target density within the annulus at radius $R$ of width $\Delta R \sim 0.06$.}
    \label{fig:BSmask} 
\end{figure*}

The masking radius around bright stars has been reduced by a factor of two in DR9 compared to DR8. Fig.~\ref{fig:BSmask} shows the stacked average density of BGS close to bright stars in \BASS/\MzLS. Distances have been rescaled to the masking radius $R_{\rm BS}(m)$ where the DR9 masking radius is represented by the smaller black circle while the radius applied in DR8 is shown by the larger red circle. Contamination around bright stars seems to be higher for the brightest stars ($8 < m < 12$) but the density profile of BGS objects (solid red line in Fig.~\ref{fig:BSmask}) shows that the contamination outside the DR9 masking radius is negligible. Results for \DECaLS~look similar to Fig.~\ref{fig:BSmask} and we believe is not necessary to included these results. Overall, in \DECaLS and \BASS/\MzLS, the reduction in the masking radius increases targetable area, and the BGS sample increase by around $28$ objects/deg$^{2}$.

\begin{figure}
	\centering
	\includegraphics[width=\columnwidth]{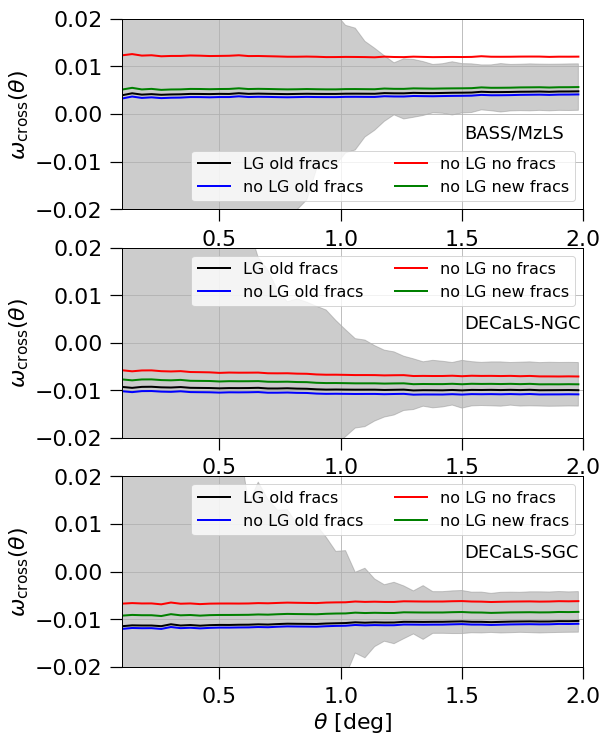}
    \caption{Angular cross-correlation function between the BGS Bright targets in each region and {\it Gaia} stars for different configurations of the BGS target selection. The {\it Gaia} stars have $12 < G < 17$. The {\it Gaia} stars have $12 < G < 17$. The shaded region shows the 1-$\sigma$ error estimated using a jackknife resampling of the data for "LG old fracs" configuration only.}
    \label{fig:xstars1} 
\end{figure}

In order to investigate the potential systematic effect associated with bright stars, we measure the angular cross-correlation between the BGS targets after masking and the stellar catalogue for BASS/MzLS, DECaLS-NGC and DECaLS=SGC. We tested several configurations for the target selection cuts: i) with and without applying the masking around large galaxies (LG), ii) considering the three options for the `FRACS cut': i) not applying the FRACS (no FRACS), ii) applying the conservative definition of DR8 (old FRACS), iii) applying a less conservative definition (new FRACS). The consequences of these different choices for the measured angular cross-correlation function of the BGS targets with {\it Gaia} stars are shown in Fig.~\ref{fig:xstars1}. As expected, the large galaxy mask has no effect on the stellar contamination and the other configurations show a negligible impact given the error bar. Therefore, we conclude  that none of these options should affect significantly the angular clustering of the BGS targets, as confirmed later in Section~\ref{sec:ang-corr}.

\begin{figure}
	\centering
	\includegraphics[width=\columnwidth]{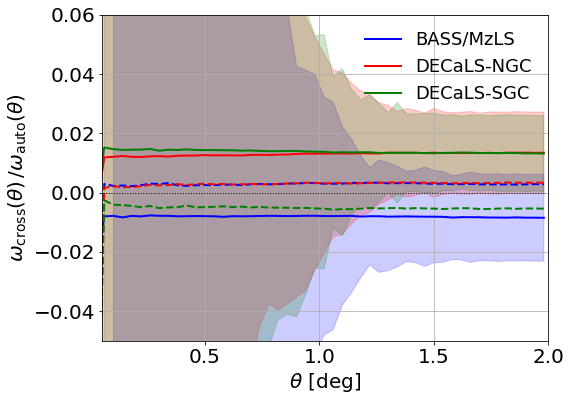}
    \caption{Ratio between the angular cross-correlation function between the BGS targets and {\it Gaia} stars and the auto-correlation function of {\it Gaia} stars. The dashed line shows this ratio after removing regions of high stellar density (stellar density < 1000 objects/\degsq). The shaded region shows the 1-$\sigma$ error estimated using a jackknife resampling of the data. The {\it Gaia} stars have $12 < G < 17$.}
    \label{fig:xstars2} 
\end{figure}

Fig.~\ref{fig:xstars2} shows the ratio between the cross-correlation function of BGS targets and stars, and the auto-correlation function of the stars. At large angular scales, this ratio gives an estimate of the fraction of stellar contamination in the BGS sample. The error bars are estimated using 100 jackknife regions for both the cross- and auto-correlation functions. The BGS targets seem uncorrelated with stars, indeed although stars represent the main systematic in the BGS selection, we note that the effect remains small compared with that seen for other DESI targets, such as the Emission Line Galaxies (ELG), which are fainter, or with Quasars (QSO), that are point-source objects. The correlation of these dark-time DESI targets with stars was shown in~\cite{Kitanidis+19}. Fig.~\ref{fig:xstars2} shows that there is no significant stellar contamination in the BGS sample, which is also consistent with the results from the star-galaxy separation when considering {\it Gaia} objects. Furthermore, the DESI Survey Validation (SV) selection will test the possibility to relax the star-galaxy criterion .

\subsection{Large galaxies}
\label{sec:xLG}

Further changes in the \BGS \, selection include not masking around the Large Galaxies of the Siena Galaxy Atlas (SGA, Moustakas, Lang, et al. in preparation), and a revisiting of the quality cuts (\QCs). The SGA galaxy catalogue has gone through a series of improvements and \TRACTOR~was run separately in regions within the confines of these galaxies, which has led to a reduction in the number of spurious objects around these galaxies in DR9 compared to DR8. These spurious objects were due to large galaxies not being appropriately fitted by TRACTOR and as a result, these galaxies were fragmented in many fake sources. In addition to that, in DR8, \TRACTOR~forced \PSF~fits to all the objects around large galaxies, compromising the photometry of the potential \BGS~targets in these regions and forcing us to mask them. In DR9, we have visually inspected around $1$ per cent of the \BGS~within the SGA mask and found $\sim 50$ per cent of them are galaxies and the remained $\sim 50$ are either stars or fragmented galaxies. We have decided to target all these to ensure completeness for clustering studies. We can reject spurious objects at a later stage. Using \GAMA~DR4, we match \BGS~with \GAMA~and for the matched objects within the \LG~mask, we were able to identify that around $40$ per cent are spectroscopically confirmed galaxies that would otherwise be rejected by the \LG mask. The comparison with \GAMA~shows that the \QCs~applied in DR8 erroneously rejected some sources that are galaxies: $80$ per cent of the sources removed by the \FRACIN, QC turn out to be galaxies, along with $50$ per cent of the objects removed by the \FRACMASKED, QC and $20$ per cent of objects in the case of \FRACFLUX.  

 \begin{figure}
	\centering
	\includegraphics[width=\columnwidth]{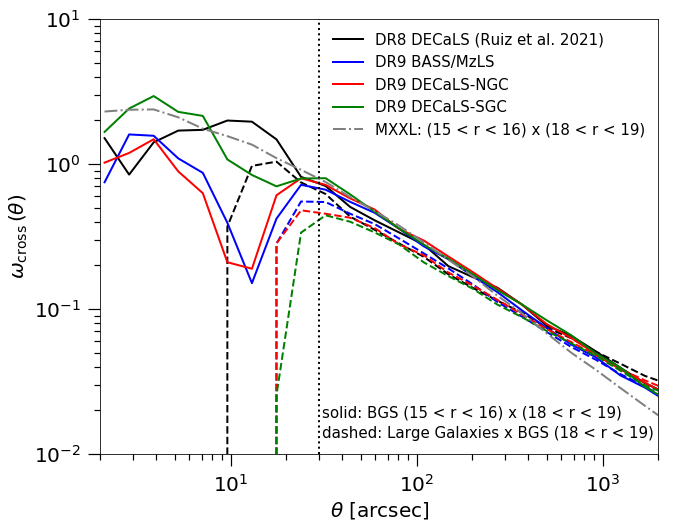}
	\includegraphics[width=\columnwidth]{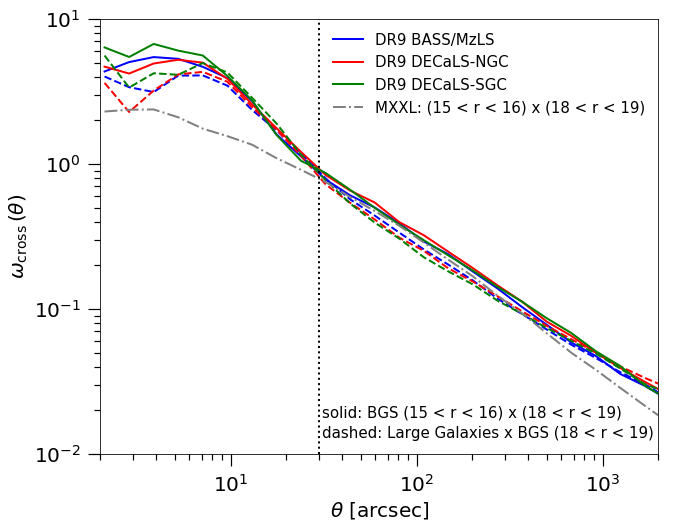}
    \caption{The angular cross-correlation function measured between faint \BGS targets in $18 < r < 19$ and  large galaxies from the \LSLGA (dashdotted) and between the same faint BGS targets and brighter \BGS targets in $15< r < 16$ (solid), the magnitude range in which most of the large galaxies reside. We also compare with the angular cross-correlation between these two bins in apparent magnitude measured in the MXXL lightcone (dashed). The vertical dotted line shows the mean LG mask radius which is about 20 arcsec. Top: with applying the large galaxy mask when selecting the BGS targets. Bottom: without applying the large galaxy mask.}
    \label{fig:xLG} 
\end{figure}

\begin{figure}
	\centering
	\includegraphics[width=\columnwidth]{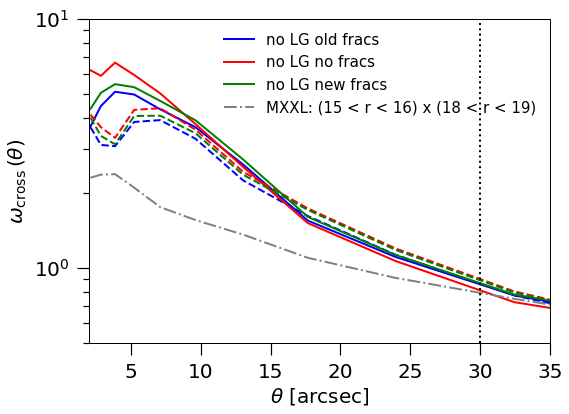}
    \caption{Same as in Fig.~\ref{fig:xLG} but here we focus on scales below the masking radius around large galaxies and we show the impact of different quality cuts (fracs). These quality cuts aim at removing spurious objects in the vicinity of the large galaxies without removing true BGS targets.} 
    \label{fig:xLG-2} 
\end{figure}

Large galaxies correspond to the brightest BGS galaxies in our sample. By comparing the angular cross-correlation between the large galaxies and BGS faint targets (with $18 < r < 19$ for instance) and the angular cross-correlation between the BGS bright targets ($15 < r < 16$ for instance) with the same BGS faint galaxies, we can estimate whether we have an excess or deficit of BGS targets in the vicinity of the large galaxies due to spurious or mis-classified sources.  A similar test for \DECaLS~DR8 was performed in~\cite{Ruiz+21} and those results are shown in black on the top panel of Fig.~\ref{fig:xLG}. This panel also shows the result of this test when masking around the large galaxies in DR9, which also includes BASS/MzLS. In DR9, the median large galaxy masking radius is about 30 arcsec (shown by the dotted vertical line in Fig.~\ref{fig:xLG}) which is twice the size of that used in DR8 as one can see from the figure where the black curve drops at smaller scale. As expected, and as we found in DECaLS DR8, when masking is applied the dashed curves that correspond to the cross-correlation function between the large galaxies and the BGS faint targets drop dramatically on scales below the masking radius, meaning that we are missing BGS targets on these scales. The bottom panel of Fig.~\ref{fig:xLG} shows a similar study but without applying the large galaxy mask. The solid and dashed curves now agree much better on scales below 30 arcsec, meaning that we recover the BGS targets in the vicinity of the large galaxies. However, the overall amplitude seems larger than what is obtained from the MXXL lightcone for BGS~\citep{Smith:2017tzz} when measuring the cross-correlation function between the bright and faint BGS galaxies in the simulation (grey curve). This suggests that the BGS selection contains some spurious objects in the vicinity of the large galaxies that could be removed by additional cuts. In Fig.~\ref{fig:xLG-2}, we show the impact on the cross-correlation signal at scales below the size of the masking radius around large galaxies of different choices for defining a quality cut based on FRACS. In Section~\ref{sec:changeTS_dr8_dr9}, we presented what this set of cuts corresponds to and in DR8 we adopted a conservative definition (old FRACS), with DR9 we investigated the effect of adopting a less conservative cut (new FRACS) or no cut at all (no FRACS). Not applying this cut increases the fraction of contamination around large galaxies which translates into a higher amplitude at these scales compared to MXXL and the other cases. The conservative approach adopted in DR8 seems to provide the best agreement with the results from the simulated lightcone. However, in the next section we will see that it also removes true BGS targets.
As a consequence, there is a balance to be found between keeping true BGS targets while removing the spurious objects around the large galaxies. For this reason, we have plans to visually inspect a random fraction of the BGS targets in the vicinity of the large galaxies using the LSVI web tool (see Section~\ref{sec:vi}) with the goal of determining the exact fraction of spurious objects and identifying common properties that could define more suitable cuts.

\section{Angular clustering measurements}
\label{sec:angular}

While many cosmological studies require knowledge of the three-dimensional distribution of  galaxies, the angular clustering also provides valuable information about both cosmology and the galaxy--halo connection. In Section~\ref{sec:ang-corr}, we present a detailed study of the angular correlation function that includes a comparison with theory and the MXXL lightcone of \cite{Smith:2017tzz} for BGS. In particular, we study the consistency between the BASS/MzLS and DECaLS BGS catalogues in terms of the angular correlation function (Section~\ref{sec:ang-consist}), then we analyse the clustering as a function of magnitude (Section~\ref{sec:ang-mag}) and as a function of colour (Section~\ref{sec:ang-col}). Finally, in Section~\ref{sec:ang-cic}, we investigate the higher-order statistics of the galaxy density field using counts-in-cells. 


\subsection{Angular correlation function}
\label{sec:ang-corr}

\subsubsection{Methodology}
\label{sec:ang-theory}
We measure the angular correlation function, $w(\theta)$, of the BGS targets using the estimator of \cite{LandySzalay93}:
\begin{equation}
     w_\textrm{LS}(\theta)=\frac{D_{1}D_{2}(\theta)-D_{1}R_{2}(\theta)- D_{2}R_{1}(\theta)+R_{1}R_{2}(\theta)} {R_{1}R_{2}(\theta)},
     \label{eq:LS}
\end{equation}
where $DD$, $DR$ and $RR$ are, respectively, the data-data, data-random and random-random pair counts at average separation $\theta$. The random catalogue is provided by  the Legacy Imaging Surveys\footnote{\url{https://www.legacysurvey.org/dr9/files/##random-catalogs-randoms}}.
The form given in equation~\ref{eq:LS} is for the cross-correlation of two samples. For the auto-correlation function, the labels 1 and 2 are indistinguishable and this simplifies to $w_\textrm{LS}(\theta) = (DD-2DR+RR)/RR$. We used the cross-correlation version of the estimator in Section~\ref{sec:xstars} (BGS targets and stars) and in Section~\ref{sec:xLG} (BGS targets and Large Galaxies). In Section~\ref{sec:cross}, we will cross-correlate the BGS targets with external spectroscopic data sets. 
We use the publicly available code TWOPCF\footnote{\url{https://github.com/lstothert/two\_pcf}} to compute the angular correlation function together with jackknife errors. These jackknife errors are obtained by dividing the footprint into 100 independent regions of similar area such that each region contains the same number of points in the random catalogue.

In order to characterise the clustering of the BGS targets, we compare it to theoretical predictions based on the halo model (e.g.~\citep[e.g.][]{PeacockSmith00,Seljak00,CooraySheth02}. In the current paradigm of galaxy formation, galaxies form within dark matter halos and the overall galaxy clustering can be modelled by two contributions: one contribution due to galaxy pairs within dark matter halos (the 1-halo term) and another contribution due to galaxy pairs in separate halos \citep[the 2-halo term; see, for example,][]{Benson:2000, Zheng:2005}. When combined, these two terms result in an approximate power law, with a feature corresponding to the 1-halo to 2-halo transition occurring around few $h^{-1}$Mpc, the typical virial radius of a halo, as first measured in the SDSS Main Galaxy Sample \citep{Zehavi:2004}. Then, to obtain a prediction for the observed angular clustering, $w(\theta)$, based on a model for the full three-dimensional clustering, $\xi(r)$, we can use Limber's approximation~\citep{Limber53} to project the real-space clustering into angular space, assuming a flat sky and small angular separations \citep[for a discussion of the validity of Limber's approximation, see][]{Simon:2007}:
\begin{equation}
    w(\theta) = \frac{2}{c} \int^{\infty}_{0} \hspace{-0.2cm}\textrm{d}z \, H(z) \, \left(\frac{\textrm{d}N}{\textrm{d}z} \right)^{2} \hspace{-0.2cm}\int^{\infty}_{0}\hspace{-0.3cm}\textrm{d}u \, \xi \left(r=\sqrt{u^{2} + x^{2}(z)\theta^{2}} \right),
    \label{eq:Limber}
\end{equation}
where $\textrm{d}N/\textrm{d}z$ is the normalised redshift distribution, $x(z)$ is the comoving distance to redshift $z$ and the integral takes account of the reduction or dilution of clustering due to the  chance alignments of uncorrelated galaxies at significantly different redshifts along the line of sight. This dilution effect is larger when the sample covers a wider range of redshift. 
Re-writing this following the notation in~\cite{Kitanidis+19}, with the centre-of-mass, $\Bar{r}=(r_{1}+r_{2})/2$, relative coordinates, $\Delta_{r}=r_{2}-r_{1}$ and where $f(\Bar{r}$ is the normalised radial distribution, the equation  becomes:
\begin{equation}
    w(\theta) = \int^{\infty}_{0} \textrm{d}\Bar{r} \, f(\Bar{r})^{2} \int^{\infty}_{-\infty} \textrm{d}\Delta r \, \xi(R,\Bar{r}).
    \label{eq:ang-model-v1}
\end{equation}

Previous studies showed that the observed correlation function can be modelled as a single power law in $r$ and $z$ up to separations of about $\simeq 10 h^{-1}$Mpc~(e.g. \citealt{Davis+83,Gaztanaga95,Maddox+96}):
\begin{equation}
    \xi(r) = \left(\frac{r_{0}}{r} \right)^{\gamma} (1+z)^{-(3+\epsilon)},
    \label{eq:power-law}
\end{equation}
where $r_{0}$ is the clustering length, the scale at which $\xi=1$, and $\gamma$ is the power-law slope. When the clustering properties do not evolve with proper coordinates, we have $\epsilon=0$~\citep{Gaztanaga95}. Assuming this power-law form for the correlation function, equation.~\ref{eq:ang-model-v1} becomes:
\begin{equation}
    w(\theta) = \theta^{1-\gamma} \, r_{0}^{\gamma} \, \sqrt{\pi} \, \frac{\Gamma(\gamma /2 - 1/2)}{\Gamma(\gamma/2)} \int^{\infty}_{0} \textrm{d}\Bar{r} \, f(\Bar{r})^{2} \, (1+z)^{(\gamma - 3)} \, \Bar{r}^{1-\gamma},
    \label{eq:ang-model-v2}
\end{equation}
This final equation can be considered as $w(\theta) = A_{\gamma, r_{0}} \theta^{1-\gamma}$, where the integral and $\Gamma$ functions have been absorbed into a constant, $A_{\gamma, r_{0}}$, whose value is set by the choices for $\gamma$ and $r_{0}$. 
Plotting $w(\theta) \times \theta^{-(1-\gamma)}$ will result in a constant if the power law model is a good description of the measured angular clustering. As one can see, there is a degeneracy between the inherent clustering amplitude and the redshift distribution of the galaxies in the sample. In what follows, we will fit the observed angular clustering with this theoretical prediction in order to extract the clustering length $r_{0}$ and slope $\gamma$, using the $\textrm{d}N/\textrm{d}z$ from the MXXL lightcone simulation~\citep{Smith:2017tzz} which matches the expected BGS redshift distribution. We note that the values of these functions and parameters that describe the BGS clustering properties could be used to create more realistic mock catalogues.


\subsubsection{Consistency between BASS/MzLS and DECaLS}
\label{sec:ang-consist}
\begin{figure}
	\centering
	\includegraphics[width=\columnwidth]{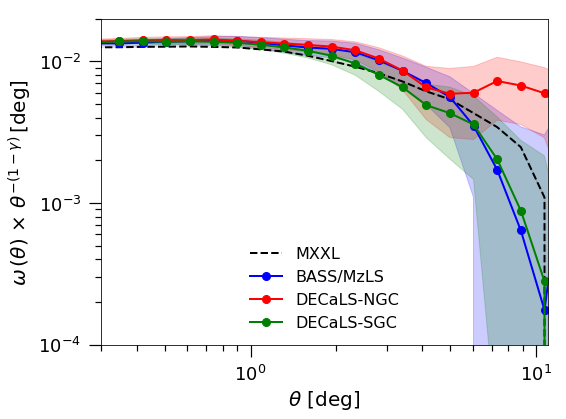}
    \vskip -0.2cm
    \caption{Angular clustering of the BGS targets in BASS/MzLS (blue), DECaLS-NGC (red), DECaLS-SGC (green), together with the results from the MXXL BGS lightcone (black). We have scaled the angular correlation function by $\theta^{-(1-\gamma)}$ to highlight departures from the power law recovered at small angular separations. The shading shows the 1-$\sigma$ error estimated using a jackknife resampling of the data.}
    \label{fig:wtheta} 
\end{figure}

First, we test the consistency between clustering in the three imaging surveys when considering BGS Bright. Fig.~\ref{fig:wtheta} shows the angular correlation function of the BGS targets in BASS/MzLS (blue), in DECaLS-NGC (blue) and in DECaLS-SGC (green), together with the angular clustering from the MXXL BGS lightcone (black). Given the choice of quantity plotted on the vertical axis, the plateau we see up to angular scales of $\simeq$~2 deg shows that the power-law form is an excellent description on these scales with $\gamma \simeq$~1.8. Beyond $\simeq$~2 deg there is a rapid reduction in the clustering away from the small-scale power law. 
We can see a very good agreement overall between the three imaging surveys and the MXXL lightcone, which is further confirmed by the results of the fitting given in Table~\ref{tab:fit-wtheta}. We find a consistent clustering length and slope between the three imaging surveys and the MXXL. Comparison with previous measurements using SDSS EDR~\citep{Stoughton+02a}, SDSS DR7~\citep{Wang:2013noa},  and the APM~\citep{Maddox+90} indicate that the angular clustering of the DESI BGS sample has a steeper  slope (i.e. the clustering strength drops more rapidly with increasing angular separation) which can be explained by the BGS redshift distribution which extends to higher redshifts than previous surveys, with a mean redshift of $z=0.2$. 

\begin{table}
    \caption{Best-fitting values for the clustering length, $r_{0}$, and power-law slope, $\gamma$, of the BGS targets in BASS/MzLS, DECaLS-NGC and DECaLS-SGC, compared to the results from the MXXL lightcone simulation, when using a power-law approximation over the fitting range $0.001 < \theta < 1$ deg.}
    \begin{tabular}{|c|c|c|}
    \hline
    dataset & $r_{0}$ [$h^{-1}$Mpc] & $\gamma$ \\
    \hline
    \multicolumn{3}{c}{BGS Bright} \\
    \hline
    BASS/MzLS     & 5.477 $\pm$ 0.117 & 1.792 $\pm$ 0.007 \\
    DECaLS-NGC    & 5.653 $\pm$ 0.118 & 1.781 $\pm$ 0.007 \\
    DECaLS-SGC    & 5.010 $\pm$ 0.079 & 1.818 $\pm$ 0.005 \\ 
    MXXL          & 4.817 $\pm$ 0.106 & 1.789 $\pm$ 0.006 \\
    \hline
    \multicolumn{3}{c}{$15 < r_\textrm{mag} < 16$} \\
    \hline
    BASS/MzLS   & 6.173 $\pm$ 0.703 & 1.642 $\pm$ 0.033 \\
    DECaLS-NGC  & 4.413 $\pm$ 0.498 & 1.761 $\pm$ 0.036 \\
    DECaLS-SGC  & 5.446 $\pm$ 0.558 & 1.698 $\pm$ 0.034 \\ 
    MXXL        & 5.731 $\pm$ 0.628 & 1.736 $\pm$ 0.039 \\
    \hline
    \multicolumn{3}{c}{$16 < r_\textrm{mag} < 17$} \\
    \hline
    BASS/MzLS   & 5.889 $\pm$ 0.359 & 1.744 $\pm$ 0.021 \\ 
    DECaLS-NGC  & 5.309 $\pm$ 0.448 & 1.761 $\pm$ 0.027 \\ 
    DECaLS-SGC  & 5.962 $\pm$ 0.368 & 1.715 $\pm$ 0.022 \\
    MXXL        & 6.189 $\pm$ 0.181 & 1.753 $\pm$ 0.029 \\  
    \hline
    \multicolumn{3}{c}{$17 < r_\textrm{mag} < 18$} \\
    \hline
    BASS/MzLS   & 5.844 $\pm$ 0.198 & 1.776 $\pm$ 0.012 \\
    DECaLS-NGC  & 6.226 $\pm$ 0.275 & 1.746 $\pm$ 0.015 \\ 
    DECaLS-SGC  & 5.514 $\pm$ 0.225 & 1.793 $\pm$ 0.015 \\ 
    MXXL        & 5.909 $\pm$ 0.206 & 1.788 $\pm$ 0.012 \\
    \hline
    \multicolumn{3}{c}{$18 < r_\textrm{mag} < 19$} \\
    \hline
    BASS/MzLS   & 5.360 $\pm$ 0.146 & 1.750 $\pm$ 0.008 \\ 
    DECaLS-NGC  & 5.444 $\pm$ 0.237 & 1.742 $\pm$ 0.013 \\ 
    DECaLS-SGC  & 5.393 $\pm$ 0.122 & 1.745 $\pm$ 0.007 \\ 
    MXXL        & 4.590 $\pm$ 0.140 & 1.803 $\pm$ 0.007 \\ 
    \hline
    \multicolumn{3}{c}{$19 < r_\textrm{mag} < 20$} \\
    \hline
    BASS/MzLS   & 5.286 $\pm$ 0.098 & 1.725 $\pm$ 0.006 \\ 
    DECaLS-NGC  & 5.336 $\pm$ 0.122 & 1.720 $\pm$ 0.007 \\ 
    DECaLS-SGC  & 5.032 $\pm$ 0.100 & 1.740 $\pm$ 0.006 \\
    MXXL        & 4.382 $\pm$ 0.107 & 1.774 $\pm$ 0.006 \\ 
    \label{tab:fit-wtheta}
    \end{tabular}
\end{table} 

\begin{figure}
	\centering
	\includegraphics[width=\columnwidth]{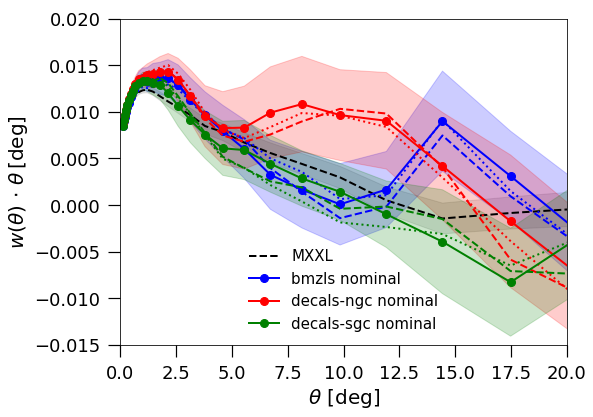}
    \vskip -0.2cm
    \caption{Angular clustering of the BGS targets when removing regions of high stellar density (dashed) or low Galactic latitude (dash-dot) compared to the original case (solid); these three estimates are consistent within the 1-$\sigma$ jackknife errors shown by the shaded regions. The angular clustering at large scales is also shown for the MXXL BGS lightcone (dashed black). Note in this plot the angular correlation function is plotted multiplied by $\theta$.}
    \label{fig:wtheta-large-scales} 
\end{figure}

In order to investigate the impact of any potential remaining imaging systematics, we also look at the angular correlation function on large scales. Fig.~\ref{fig:wtheta-large-scales} shows the angular clustering up to 20 deg for the three imaging surveys and the MXXL. The solid curves correspond to the nominal configuration, the dashed ones to the case where we remove regions of high stellar density (i.e. we keep stardens < 1000/\degsq), and the dotted curves to the case where we remove regions of low Galactic latitude (we keep |b| > 30 deg). These two tests have a negligible impact on the clustering given the size of the error bars at these large scales. The overall agreement is reasonably good. At angular scales between 5 and 15 deg, DECaLS-NGC seems to have a higher amplitude but again, the errors bars are important at these very large angular scales. One may question the validity of the jackknife errors at these scales. In order to test this we computed the error bars using 10, 25 and 50 jackknife regions and compared with the errors when using 100 jackknife regions. We notice a slight under-estimation when increasing the size of the jackknife region as expected, but otherwise the effect remains small which validates our interpretation of Fig.~\ref{fig:wtheta-large-scales}: the difference in clustering amplitude in this regime is consistent with being due to a statistical fluctuation.

\subsubsection{Clustering as a function of magnitude}
\label{sec:ang-mag}

\begin{figure}
	\centering
	\includegraphics[width=\columnwidth]{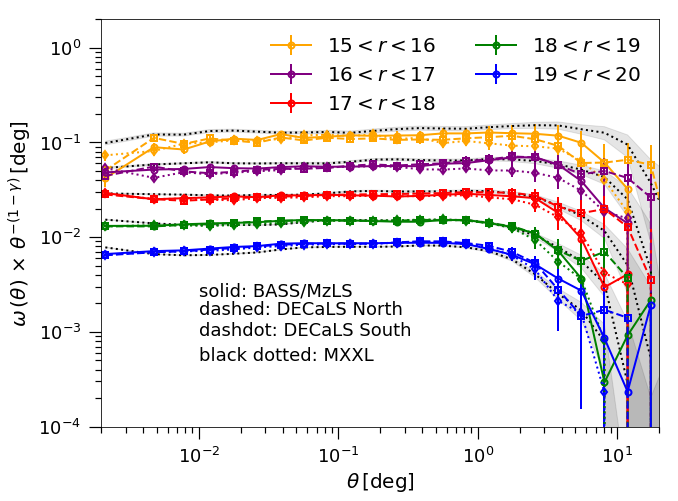}
    \vskip -0.2cm
    \caption{Angular clustering as a function of the apparent magnitude in the r-band for BASS/MzLS (solid), DECaLS-NGC (dashed), DECaLS-SGC (dashdot), together with the results from the MXXL BGS lightcone mock (dotted)}.
    \label{fig:wtheta-rmag} 
\end{figure}

As an additional check for systematics, we compute the angular correlation function for different apparent magnitude bins and compare the results of the BGS targets with the MXXL simulation as shown in Fig.~\ref{fig:wtheta-rmag}. The quantity plotted on the y-axis, $w(\theta)\times \theta^{-(1-\gamma)}$, was choosen such that one can see the domain of validity of the power-law form, as for Fig.~\ref{fig:wtheta}. Table~\ref{tab:fit-wtheta} presents the results of the power-law fitting on both DR9 and MXXL for the five apparent magnitude bins we consider.


\begin{figure}
	\centering
	\includegraphics[width=\columnwidth]{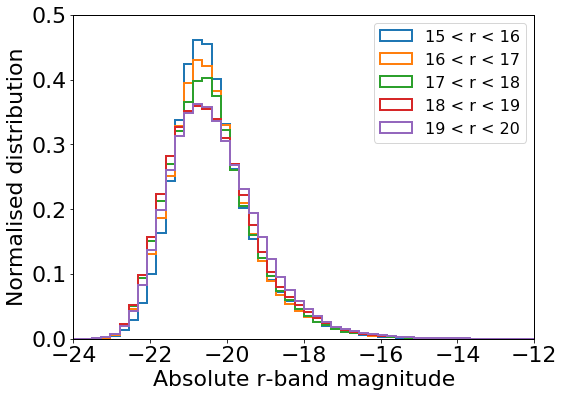}
	\includegraphics[width=\columnwidth]{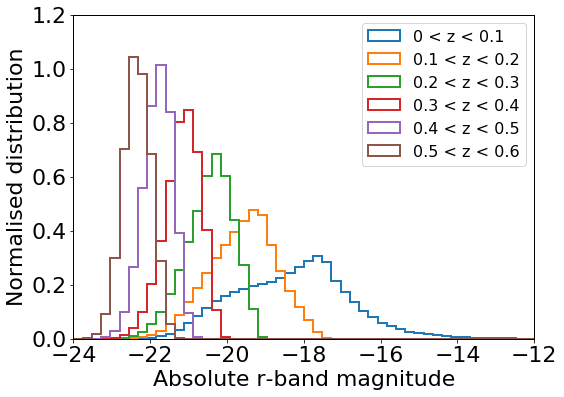}
	\includegraphics[width=\columnwidth]{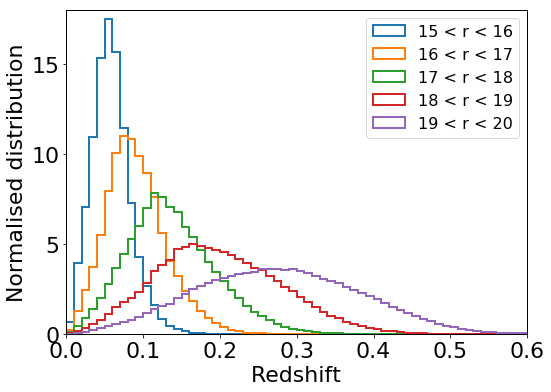}
    \caption{Top: Normalised absolute $r$-band magnitude distribution in MXXL for different apparent $r$-band magnitude slices. Middle: Normalised absolute $r$-band magnitude distribution in MXXL for different redshift slices. Bottom: Normalised redshift distribution in MXXL for different apparent $r$-band magnitude slices.}
    \label{fig:mxxl} 
\end{figure}

In order to help interpret these results we first quantify some properties of the matching MXXL mock catalogue. In Fig.~\ref{fig:mxxl}, we see that the distribution of absolute magnitude has very little dependence on the apparent magnitude range of the sample. Hence we would expect each of our apparent magnitude samples to be dominated by galaxies of the same absolute magnitude and hence have similar 3-dimensional clustering, $\xi(r)$. The main way in which the samples differ in their normalized $dN/dz$ shown in the bottom panel of Fig.~\ref{fig:mxxl}. The shallower more sharply peaked $dN/dz$ of the brighter samples will lead to stronger angular cluster, due to the $(dN/dz)^2$ term in Limber's equation (equation.~\ref{eq:Limber}), and the break away from the small scale power-law occur on larger angular scales dues to a fixed comoving separation subtending a larger angle at low redshift. This is precisely how the observational results shown in Fig.~\ref{fig:wtheta-rmag} behave.

To summarise, we find an overall consistent clustering strength and slope between the three imaging surveys and MXXL. Moreover, compared to the reference SDSS measurements~\citep{Wang:2013noa}, the DESI BGS allows us to obtain more precise measurements due to the larger size of the sample and greater reliability on large scales.



\subsubsection{Clustering as a function of colour}
\label{sec:ang-col}

A galaxy's colour reflects its composite stellar population, which in turn depends on its star formation history, the chemical enrichment history of the star-forming gas, and the attenuation of the starlight by dust; these processes are influenced by the mass of the galaxy's host dark matter halo (for reviews, see \citealt{Conroy:2013,SomervilleDave15}). Therefore, massive galaxies with red colours typically have older stellar populations while galaxies with intermediate masses are bluer and younger with higher star formation rates. 

\begin{figure}
	\centering
	\includegraphics[width=\columnwidth]{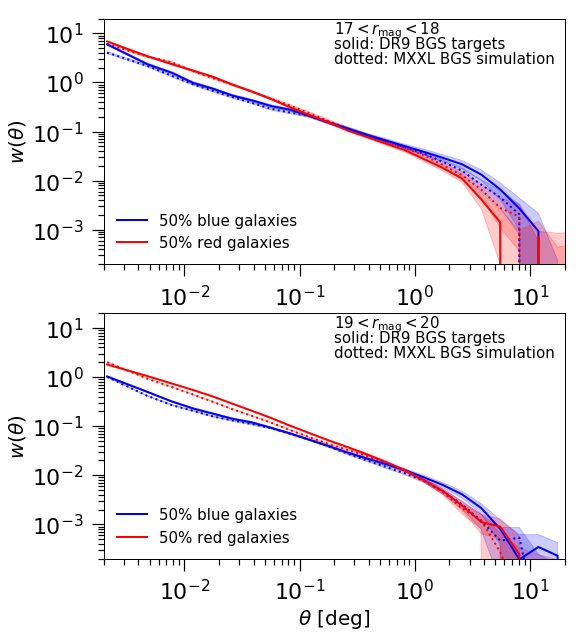}
    \caption{Angular clustering of the BGS targets with $17 < r_\textrm{mag} < 18$ (top  panel) and $19 < r_\textrm{mag} < 20$ (bottom panel), for samples divided by colour into red and blue galaxies. The BGS measurements are shown by solid lines. The results using the MXXL BGS lightcone mock are also shown (dotted) for the same configurations as for the data.}
    \label{fig:wtheta-colour} 
\end{figure}

In order to disentangle the colour, luminosity and redshift dependence of the galaxy  clustering, we compute the colour-dependent clustering in two apparent magnitude bins for both BGS DR9 and MXXL. For each apparent magnitude bin, we split the sample into the 50 per cent bluest galaxies and 50 per cent reddest galaxies using $g-r$ colour. We found that considering a fixed fraction of blue/red galaxies instead of fixed colour cuts results in a fairer comparison between BGS DR9 and MXXL, as the colour distribution in the MXXL simulation does not match perfectly that of the observations, particularly at fainter magnitudes. The results of this exercise are shown in Fig.~\ref{fig:wtheta-colour}, where the top panel corresponds to galaxies with $17 < r_\textrm{mag} < 18$ and the bottom panel to galaxies with $19 < r_\textrm{mag} < 20$. For each magnitude bin the angular clustering of blue and red galaxies is shown for DR9 BGS (solid) and the MXXL simulation (dotted). As expected, we can see that red galaxies are more strongly clustered than blue ones at intermediate to small angular separations. The overall agreement with the lightcone is good over a large range of angular scales, which thus validates the colour-assignment procedure in MXXL.

\subsection{Higher-order statistics using counts-in-cells}
\label{sec:ang-cic}

If the density field is a purely Gaussian random field, then its probability distribution function can be described by just two numbers: the mean and the variance. A Gaussian primordial density field is well supported by observations of the CMB~\citep{Planck18}. However, one can easily show that, in the gravitational instability scenario, this primordial distribution of density fluctuations will evolve into a distinctly asymmetric density field. Thus, the observed higher-order moments of the local density field contain information besides the two-point statistics, such as the departure from Gaussianity, which can inform us about the growth of cosmic structures, and more specifically on the bias between galaxies and the underlying matter distribution (see the review by \citealt{Bernardeau:2002}). Moreover, in order to produce more realistic mock catalogues, it is essential to reproduce the higher-order clustering statistics of the BGS sample, especially for regions of high density where spectroscopic incompleteness due to the finite size of the fibre allocation\footnote{The DESI patrol radius is about $1.4~\arcmin$ which corresponds to 0.017~deg.} has a significant impact on clustering~\citep{Burden+17,Hahn+17,Bianchi+18,Smith:2019,BianchiVerde20}. These higher-order statistics can be explored using counts-in-cells (CIC, see for example \citealt{White79,Peebles80, Fry:1993}).

The CIC analysis of projected galaxy counts in wide-field galaxy surveys has a long history, stretching back to visually measured counts on photographic plates \citep{Groth:1977}. \cite{Gaztanaga:1994} measured the distribution of CIC up to ninth order from the Automated Plate Machine survey \citep{Maddox+90}, showing that the galaxies are essentially unbiased tracers of the matter distribution on large scales. \cite{Ross+06,Ross+07} applied CIC to the third release of SDSS in order to measure the higher-order angular correlation functions of SDSS that can be used for testing the hierarchical clustering model and higher-order bias terms. \cite{Salvador+19} developed the technique to measuring the linear and non-linear galaxy bias of the Dark Energy Survey Science Verification data. More recently, \cite{ReppSzapudi20} developed a theoretical prediction of the CIC galaxy probability function as a function of $\sigma_{8}$ and $b$ to measure these parameters from the SDSS Main Galaxy Sample.

\begin{figure}
	\centering
	\includegraphics[width=0.8\columnwidth]{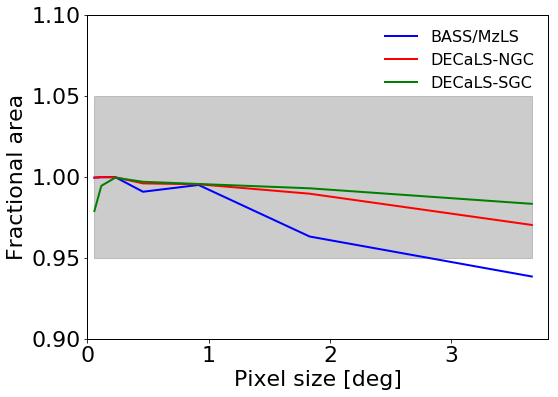}
    \vskip -0.2cm
    \caption{Ratio of the effective area after and before removing the outliers in the HEALPix distribution based on the random catalogue for the three imaging surveys: BASS/MzLS (blue), DECaLS-NGC (red), DECaLS-SGC (green). The shaded grey region shows 10\% variation of this fractional area.}
    \label{fig:fracarea} 
\end{figure}

With the goal of providing a complete characterisation of the clustering properties of the BGS sample, in this section we investigate the higher-order statistics of the density field up to fourth order: mean, variance, skewness and kurtosis by making use of the CIC method. We use the HEALPix\footnote{https://healpix.sourceforge.io} package~\citep{Gorski+05} which divides the sky such that each pixel covers the same surface area. This method works for the entire DESI footprint, unlike the one used in \cite{Kitanidis+19}, which is based on a transformation of the angular coordinates into cartesian coordinates. This is a good approximation for regions close to the Galactic plane, such as the rectangles defined in their paper for DECaLS, but is no longer valid when considering BASS/MzLS for instance. In the HEALPix pixelation, the lowest resolution partition is comprised of 12 base pixels and the resolution increases by dividing each pixel into four new ones such that $\textrm{nside} = 2^\textrm{resolution}$ is the number of pixels per side and $\textrm{npix} = 12 \times \textrm{nside}^2$ is the total number of pixels in the map. In what follows, we consider resolutions above 4 to the maximum 10. The maximum resolution corresponds to a cell size of roughly 0.06 deg which is above the DESI patrol radius. For each resolution of the HEALPix maps, we remove pixels that are not fully within the survey boundaries by determining a threshold based on the expected number density using the random catalogue. The threshold is determined such that these outliers in the HEALPix pixels distribution are removed while decreasing the effective area by less than 10\%, as confirmed in Fig.~\ref{fig:fracarea} which shows the difference in effective area after and before removing the outliers for BASS/MzLS (red), DECaLS-NGC (blue) and DECaLS-SGC (red) based on the random catalogue. 

\begin{figure*}
	\centering
	\includegraphics[scale=0.3]{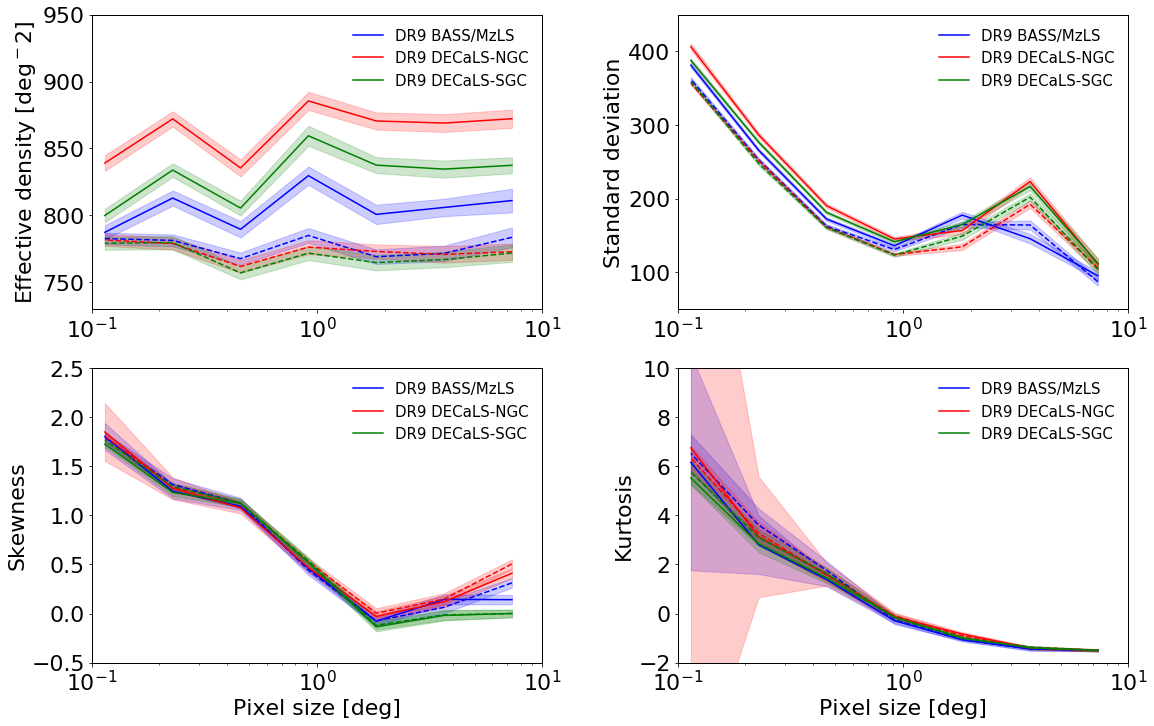}
    \vskip -0.2cm
    \caption{Effective mean density, standard deviation, skewness and kurtosis as a function of HEALPix cell size in degrees for the MXXL lightcone (dashed) and for the BGS DR9 targets (solid) where both are restricted to the same imaging region with errors bars from 100 jackknife regions.}
    \label{fig:cic} 
\end{figure*}

For each resolution of the HEALPix map, we compute the effective mean density per square degree, the standard deviation, the skewness and the kurtosis. 
\cite{SzapudiColombi96} showed that the CIC statistics are sensitive to sample variance (shot noise, edge effects and finite volume) and to measurements errors due to the finite number of sampling cells. \cite{Szapudi98} proposed a method of infinite oversampling that enables the noise that is introduced by having only one set of sampling cells to be beaten down and thus to eliminate the measurement errors. In order to reproduce this oversampling effect, we dither by a fraction of cell size each HEALPix map, compute the CIC statistics for each rotation and take the average. In practise, first we convert the RA, $\it {Dec}$ into $x,y,z$ coordinates and then rotate the coordinates by an angle $\phi$ (in degrees) around an arbitrary rotation axis vector. The angle $\phi$ is randomly chosen in a Gaussian distribution of width the HEALPix cell size (we also tried twice and five times the HEALPix cell size). Eventually we convert back the shifted $x,y,z$ into new RA, $\it {Dec}$. We do 5 rotations and compute the mean and standard deviation of each quantity above. We did not find any shift in the mean value of each CIC statistics associated with this shifting of pixels, which confirms that we are carrying out a robust sampling and that the tails of the counts distribution are well measured and not unduly affected by the sampling of extreme voids or overdensities. 
In order to estimate errors, we define a set of 100 jackknife regions, the same set for every pixel size, and we compute the effective density, standard deviation, skewness and kurtosis in each region, take the mean and the standard deviation. First, we test the procedure using the MXXL lightcone that we split into BASS/MzLS, DECaLS-NGC and DECaLS-SGC regions. Fig.~\ref{fig:cic} shows the results of the MXXL lightcone in dashed for each statistic as a function of HEALPix cell size for BASS/MzLS (blue), DECaLS-NGC (red), DECaLS-SGC (green) with the coloured regions representing the 1-$\sigma$ errors from the 100 jackknife regions. As expected, the measurements for the different MXXL regions all agree to within the errors. The solid curves show the same results for the BGS DR9 targets with their jackknife errors. The values of the target density for the three imaging surveys are consistent with the ones given in Section~\ref{sec:comp_dr8} with 7\% difference at maximum when correcting for the magnitude and colour shift between BASS/MzLS and DECaLS. The other statistics show a better agreement between the three imaging regions of the BGS data and with the MXXL lightcone, even for the third (skewness) and fourth (kurtosis) moments of the galaxy density field although no direct information about these higher-order statistics was included in the creation of the MXXL lightcone for BGS. We note that both the skewness and kurtosis are non-zero as expected from a primordial Gaussian random density field which evolved under gravitational instability and led to the hierarchy of gravitationally-bound structures that form the cosmic web with filaments, sheets, knots and voids.


\section{Cross-correlations with external spectroscopic data}
\label{sec:cross}

In this section, we study the cross-correlations between the BGS targets and external spectroscopic data sets that we present in Section~\ref{sec:ext-cat} in order to determine the real-space projected cross-correlation function $w_{p}(r_{p})$ in Section~\ref{sec:wp} and the redshift distribution ${\rm d} N/{\rm d}z$ of the BGS targets in Section~\ref{sec:dndz}.

\subsection{External spectroscopic data sets}
\label{sec:ext-cat}
\begin{figure}
	\centering
	\includegraphics[width=\columnwidth]{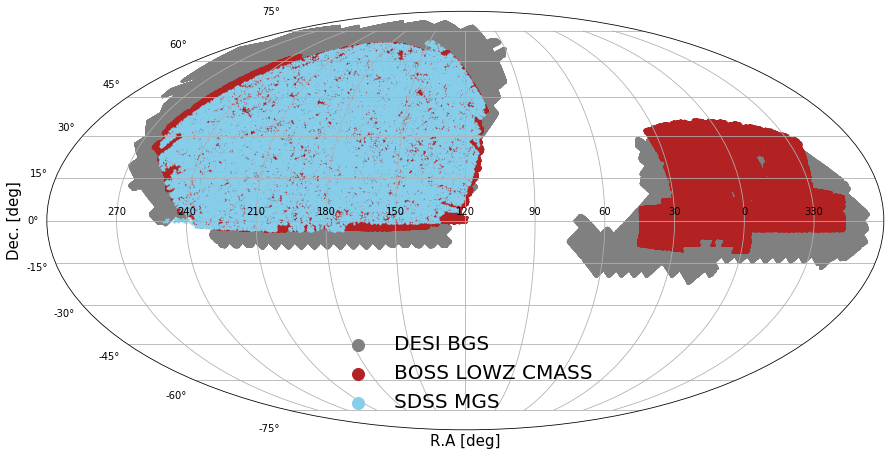}
    \vskip -0.2cm
    \caption{Footprint of the different datasets used in the  analysis in \S~5: in grey the BGS using DR9 DESI Legacy Imaging Surveys, in red SDSS MGS and in blue BOSS LOWZ and CMASS.}
    \label{fig:footprint} 
\end{figure}

In order to study the clustering properties of the BGS targets as a function of redshift, we make use of two external spectroscopic data sets that overlap both in area and redshift with BGS. At $z < 0.2$, in the NGC, we use the SDSS DR7 MGS clustering catalogue~\citep{Ross+15} that covers 6813 deg$^{2}$ and contains 63,163 SDSS DR7 spectroscopically identified galaxies. This dataset overlaps with about 915 deg$^{2}$ of BASS/MzLS and with about 1410 deg$^{2}$ of DECaLS-NGC.
In both the NGC and SGC, up to $z < 0.6$, we use the LOWZ and CMASS clustering catalogues~\citep{Reid+16} from  DR12 of SDSS-III BOSS~\citep{Eisenstein+11, Dawson+13}. The LOWZ sample contains galaxies at $z \leq 0.4$ covering 8,337 deg$^{2}$ and the CMASS sample contains galaxies in $0.4 \leq z \leq 0.7$ covering 9,376 deg$^{2}$. In total, the overlap with BASS/MzLS corresponds to roughly 2965 deg$^{2}$, the one with DECaLS-NGC to 4335 deg$^{2}$ and the one with DECaLS-SGC to 2585 deg$^{2}$.
Fig.~\ref{fig:footprint} shows the footprint of the DR9 Legacy Imaging Surveys restricted to the DESI footprint, together with the footprint of SDSS MGS and BOSS LOWZ-CMASS.

\subsection{Clustering as a function of redshift}
\label{sec:wp}
By making use of cross-correlation with samples of known redshift, we can infer the projected real-space cross-correlation function, $w_{p}(r_{p})$, which corresponds to the integral of the 2D spatial correlation function $\xi(r_{\parallel},r_{\perp})$ over the line of sight direction $r_{\parallel}$ \citep{Peebles80}. This projected real-space correlation function can be related to the angular correlation function $w(\theta)$. Following \cite{Padmanabhan+09}, if we consider that the spectroscopic sample lies at $\chi = \chi_{0}$, then according to the flat-sky approximation and by applying the Limber approximation, the relation simplifies to:
\begin{equation}
    w(\theta) = f(\chi_{0}) w_{p}(r_{p}),
    \label{eq:ang-proj}
\end{equation}
where $f(\chi_{0})$ is the normalised radial distribution of the photometric sample. We can generalise this to a narrow spectroscopic redshift slice where we consider the clustering as constant over the redshift slice and we can assume that the redshift of the photometric galaxy is the same as the one of the spectroscopic galaxy it is correlated with. It allows us to rebin the pair counts of the angular correlation function in transverse separation, $w_{\theta}(r_{p})$. We can thus rewrite equation~\ref{eq:ang-proj} as:
\begin{equation}
   w_{\theta}(r_{p}) =  \langle f(\chi) \rangle w_{p}(r_{p}),
\end{equation}
where $f(\chi)$ is the normalised radial distribution of the photometric sample over the spectroscopic bin considered. In what follows, we use the MXXL lightcone simulation to estimate the contribution from the redshift distribution of the BGS targets. Recently, this observable, the projected cross-correlation function binned into transverse comoving  radius, was also developed for improving the measurement of the Baryon Acoustic Oscillations feature for a sparse spectroscopic sample \citep{Patej+18}. The first detailed application to data was proposed in \cite{Zarrouk+21} where a sample of sparse quasars from eBOSS was cross-correlated with a dense photometric sample of galaxies using the DR8 DESI Legacy Imaging Surveys. 

We use the Landy-Szalay estimator as defined by equation~\ref{eq:LS} where we count pairs in 16 logarithmically spaced angular bins between $\theta$=0.001 deg and $\theta$=1 deg.
We divide the external spectroscopic datasets into redshift slices of width $\Delta z =0.1$, and each redshift slice is cross-correlated with the BGS catalogue. Errors bars are estimated using the bootstrap technique presented in \cite{Kitanidis+19}, using 500 bootstrap realisations. We checked that our results are robust to the choice of the number of bootstrap realisations. We can also compare the projected real-space cross-correlation function with theory by assuming a power-law  model for the correlation function, which gives:
\begin{equation}
    w_{p}(r_{p}) = r_{p}^{1-\gamma} r_{0}^{\gamma} \sqrt{\pi} \frac{\Gamma(\gamma /2 - 1/2}{\Gamma(\gamma /2)}.
    \label{eq:wprp}
\end{equation} 

\begin{figure}
	\centering
	\includegraphics[width=\columnwidth]{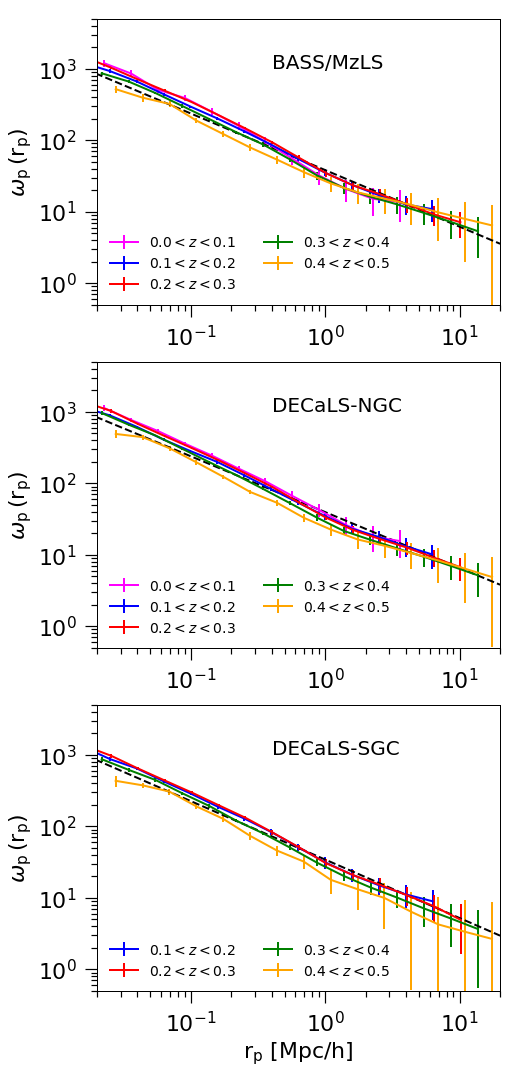}
    \vskip -0.2cm
    \caption{Real space projected cross-correlation function of the BGS targets with external spectroscopic datasets, SDSS MGS and BOSS LOWZ-CMASS for BASS/MzLS (top), DECaLS-NGC (middle) and DECaLS-SGC (bottom).}
    \label{fig:wrp} 
\end{figure}

Fig.\ref{fig:wrp} shows the real-space projected cross-correlation function of the BGS targets in BASS/MzLS (top), DECaLS-NGC (middle) and DECaLS-SGC (bottom). The colours represent the cross-correlation functions with different redshift slices of the external spectroscopic datasets (SDSS MGS DR7 for BASS/MzLS and DECaLS-NGC, BOSS LOWZ-CMASS for all three DECaLS regions). The dashed black line shows the theoretical prediction when assuming a power-law for the correlation function as given by Eqn.~\ref{eq:wprp} with the clustering length $r_{0}$ and slope $\gamma$ corresponding to the best-fitting parameters from Table~\ref{tab:fit-wtheta} for each imaging region. We thus obtain a consistency between the behaviour of the angular correlation function, $w(\theta)$, and the projected real-space cross-correlation function, $w_{p}(r_{p})$. Fig.~\ref{fig:wrp} also shows a slight decrease of the clustering amplitude with increasing redshift which is in competition between the redshift evolution of the absolute magnitude - as redshift increases, galaxies are intrinsically fainter which is confirmed in MXXL and shown in the middle panel of Fig.~\ref{fig:mxxl} - and the redshift evolution of the clustering amplitude - as redshift increases, the clustering amplitude decreases. Using the formula for the luminosity-dependence of the clustering of \cite{Norberg+01} for the extreme slices where the peak luminosity shifts from $-18$ in the lowest redshift bin and to $-22$ in the highest redshift bin, we found a $0.73$ reduction in clustering amplitude moving from high to low redshift. At the same time, the growth factor at these two redshifts changes the clustering amplitude by $1.51$, which gives an overall 10\% effect on the clustering amplitude, i.e. $1.51 \times 0.73=1.10$.

\subsection{Clustering-redshift dN/dz}
\label{sec:dndz}
The cross-correlation signal between a photometric sample and spectroscopic data sets can also be used to infer the redshift distribution of the photometric sample. Using cross-correlations to get information about the properties of the photometric galaxies is not new. For instance \cite{SeldnerPeebles79} used cross-correlations of quasars and galaxy counts to test how galaxies are distributed around quasars and to see if this depends on the redshift and apparent magnitude of the quasar. \cite{Schneider+06} first discussed the idea of using a spectroscopic sample to determine the redshift distribution of photometric galaxies, then \cite{Newman08} expanded on this and described the clustering-redshift technique. We follow the approach proposed by~\cite{Menard+13} and \cite{Rahman+15} which uses smaller scales and accounts for the redshift evolution of the galaxy bias. We measure the angular cross-correlation function between the BGS targets and narrow redshift slices of the spectrocopic datasets ($\Delta z =0.05$). For each redshift slice $\Delta z_{i}$, we compute the angular cross-correlation function $w_{ur}(\theta,\Delta z_{i})$, then we integrate over an annulus centered on the spectroscopic object from $\theta_{\rm min}$ to $\theta_{\rm max}$. Indeed, \cite{Menard+13} showed that the sensitivity of the cross-correlation signal is improved when including information from many clustering scales. In order to maximise the signal-to-noise ratio, we can weight each measurement by $\theta^{-1}$. Applying this weighting gives the following integrated cross-correlation function:
\begin{equation}
    \Bar{w}_{\rm ur}(z) = \int_{\theta_{\rm min}}^{\theta^{\rm max}} {\rm d}\theta \, \frac{w_{\rm ur}(\theta,z)}{\theta},
    \label{eq:wur}
\end{equation}
where we set $\theta_{\rm min}, \theta_{\rm max}$ to match a fixed range of projected radii $r_{p,\rm min}, r_{p,\rm max}$ so that we can probe the same range of physical scales as a function of redshift.
We use $r_{p,\rm min} = 0.05 h^{-1}$Mpc and $r_{p,\rm max} = 5 h^{-1}$Mpc.

Finally, the redshift distribution can be inferred from the integrated cross-correlation function using:
\begin{equation}
    \frac{{\rm d}N}{{\rm d}z} \propto \frac{\Bar{w}_{\rm ur}(z)}{b_{u}(z)b_{r}(z)},
    \label{eq:dndz}
\end{equation}
where $b_{u}$ is the bias of the BGS sample and $b_{r}$ is the bias of the spectroscopic sample. For SDSS MGS DR7 and BOSS LOWZ-CMASS, we use the large-scale bias measurements of~\cite{Ross+15,Howlett+15} and of~\cite{Rodriguez+16} respectively. For both spectroscopic samples the redshift evolution of the bias is negligible over the redshift slice $\Delta z = 0.05$. For the BGS targets, we use the DESI Final Design Report values~\citep{Desi16a} and we find consistency when comparing with the MXXL BGS lightcone. However, we highlight that the aim is not to provide the most accurate redshift distribution using this technique but rather to show an illustration of it and provide an additional check of the clustering evolution with redshift. Moreover, we note that the Survey Validation is currently providing accurate redshifts for building truth tables and testing the performance of the spectroscopic pipeline.

\begin{figure}
	\centering
	\includegraphics[width=\columnwidth]{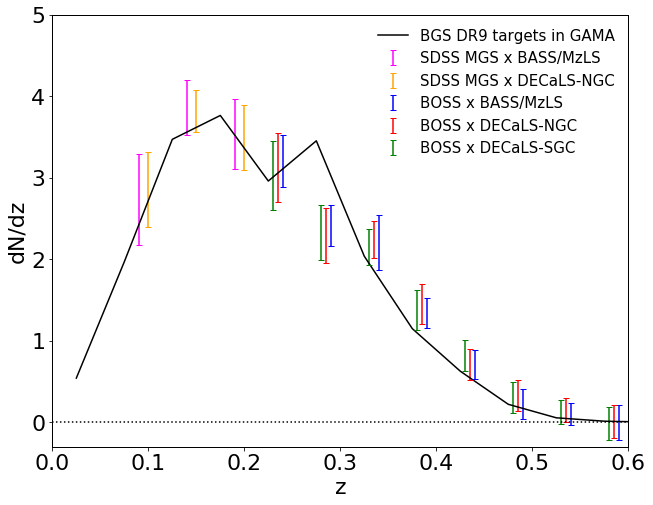}
    \vskip -0.2cm
    \caption{Normalised Redshift distribution for the BGS DR9 targets that match with GAMA galaxies (black), for the BGS DR9 targets using clustering-redshifts by cross-correlating with SDSS MGS galaxies (magenta and yellow) and BOSS LOWZ-CMASS galaxies (blue, red, green).}
    \label{fig:dndz} 
\end{figure}

Fig.~\ref{fig:dndz} shows the redshift distribution ${\rm d}N/{\rm d}z$ of the BGS targets in BASS/MzLS, DECaLS-NGC and DECaLS-SGC using this clustering-redshift technique with SDSS MGS DR7 galaxies in $0.07 < z < 0.2$ and BOSS LOWZ-CMASS galaxies $0.2 < z < 0.6$. We confirm that the inferred redshift distributions are all in agreement between the three imaging regions, and also with the redshift distribution of the BGS targets that match GAMA galaxies using their spectroscopic redshifts (in black). 

\section{Conclusions}
\label{sec:concl}
We have presented the target selection pipeline that selects the BGS bright targets ($r_{\rm mag} \leq 19.5$) from the latest release of the Legacy Imaging Surveys (DR9) that uses BASS, MzLS and DECaLS over the DESI footprint of 14,000 deg$^{2}$. This includes  several changes with respect to what was first presented using DECaLS DR8 in~\cite{Ruiz+21} that we summarize here:
\begin{itemize}
    \item Thanks to major improvements in the iterative fitting around bright stars, the radius of the bright star mask is half the size in DR9 than it was in DR8. We have checked that this change does not introduce any spurious effects. We found that our star-galaxy separation based on GAIA yields less than 2\% stellar contamination which is also confirmed by the cross-correlation signal of BGS targets and bright stars at large scales.
    \item Thanks to major improvements in the photometry, we no longer need to apply a spatial masking around large galaxies: we were missing true BGS targets in the vicinity of the large galaxies that we are now able to recover. We checked that this change improves the completeness with respect to GAMA which is highly complete in $14 < r < 19$. We also developed a visual inspection webtool which confirms that not applying the large galaxy mask does not introduce a significant fraction of spurious objects.
    \item Using this visual inspection webtool, we also found that we could make a less conservative choice on some of the quality cuts that involve FRACIN, FRACMASKED and FRACFLUX which increases the completeness slightly while ensuring a negligible fraction of spurious objects in the vicinity of the large galaxies.
    \item Finally, the DR9 selection cuts yield a completeness with respect to GAMA DR4, which is complete in $14 < r < 19$, that is above 99\%. DR9 also results in a BGS bright sample ($r \leq 19.5$) that meets the requirement for target density which is above 800 deg$^{-2}$ for the three imaging surveys.
\end{itemize}

After presenting the target selection cuts and studying the main systematics, we characterised the clustering properties of the BGS bright sample by looking at several statistics:
\begin{itemize}
    \item The angular clustering shows very good consistency between the three imaging surveys and with the MXXL BGS lightcone\citep{Smith:2017tzz}. It is also consistent with a power-law model on angular scales below 1 deg and it gives comparable clustering strength $r_{0}$ and slope $\gamma$, both between the three imaging surveys, with the mock and with previous measurements in the litterature.
    \item The angular clustering as a function of magnitude confirms that brighter samples have higher amplitude and bigger clustering strength that corresponds to brightter galaxies occupying more massive halos.
    \item The colour-dependent clustering confirms that brighter and redder galaxies are more strongly clustered. We also found a very good agreement with MXXL when splitting both samples by magnitude and fixed fraction of blue/red galaxies.
    \item We also investigated higher-order clustering properties by making use of the counts-in-cells technique using the HEALPix package. We compared the mean density, standard deviation, skewness and kurtosis of the galaxy density field in BASS/MzLS, DECaLS-NGC, DECaLS-SGC for the data and MXXL.
    \item Finally, we looked at the redshift evolution of the clustering by cross-correlating the BGS targets with external spectroscopic datasets that overlap in area and redshift. We used SDSS MGS DR7 and BOSS LOWZ-CMASS to compute the projected real-space cross-correlation function in redshift slices of width $\Delta z = 0.1$. We found very good agreement and a consistent clustering with the prediction from a power-law model using the best-fitting values for $r_{0}$ and $\gamma$ that we measured by fitting the angular clustering. We also inferred the redshift distribution of the BGS targets using the clustering-redshift technique. We were able to recover a redshift distribution that is consistent between the three imaging surveys and with the one we estimated by matching the BGS targets with GAMA galaxies and by using their spectroscopic redshifts.
\end{itemize}

As a summary, this work validates the selection for the BGS bright sample which will be very close to the final one. We have characterised and studied several clustering statistics that provide the first large-scale angular clustering analysis of the DESI BGS. Modeling the sample is an important first step for doing cosmology with DESI, and our clustering results will also aid in the creation and validation of accurate mock catalogues just as the ones we are currently developing with the AbacusSummit~\footnote{\url{https://abacussummit.readthedocs.io/en/latest/}} simulations (Maksimova et al. \textit{in prep.}) which uses the Abacus code~\citep{Garrison+18,Metchnik09}. However, additional work is required using the spectroscopic data of the ongoing Survey Validation period in order to investigate 1) the spectroscopic completeness, 2) the fiber assignment efficiency and impact on the clustering, 3) the 3D clustering. Moreover, the final target selection will also include a fainter sample that will be fully characterised in a future work. 

\section*{Acknowledgements}
PZ, OR-M, SC, PN and CB acknowledge support from the Science Technology Facilities Council through ST/P000541/1 and ST/T000244/1.
OR-M is supported by the Mexican National Council of Science and Technology (CONACyT) through grant No. 297228/440775 and funding from the European Union’s Horizon 2020 Research and Innovation Programme under the Marie Sklodowska-Curie grant agreement No 734374.
JM gratefully acknowledges support from the U.S. Department of Energy, Office of Science, Office of High Energy Physics under Award Number 
DESC0020086 and from the National Science Foundation under grant AST-1616414.
Authors want to thank the GAMA collaboration for early access to GAMA DR4 data for this work. Some of the results in this paper have been derived using the healpy and HEALPix package. We acknowledge the usage of the HyperLeda database (\url{http://leda.univ-lyon1.fr}).This work also made extensive use of the NASA Astrophysics Data System and of the astro-ph preprint archive at arXiv.org.  

This work used the DiRAC@Durham facility managed by the Institute for Computational Cosmology on behalf of the STFC DiRAC HPC Facility (www.dirac.ac.uk). The equipment was funded by BEIS capital funding via STFC capital grants ST/K00042X/1, ST/P002293/1 and ST/R002371/1, Durham University and STFC operations grant ST/R000832/1. DiRAC is part of the National e-Infrastructure. 

This research used resources of the National Energy Research Scientific Computing Center (NERSC). NERSC is a U.S. Department of Energy Office of Science
User Facility operated under Contract No. DE-AC02-05CH11231.

This research is supported by the Director, Office of Science, Office of High Energy Physics of the U.S. Department of Energy under Contract No. 
DE–AC02–05CH1123, and by the National Energy Research Scientific Computing Center, a DOE Office of Science User Facility under the same 
contract; additional support for DESI is provided by the U.S. National Science Foundation, Division of Astronomical Sciences under Contract No. 
AST-0950945 to the NSF’s National Optical-Infrared Astronomy Research Laboratory; the Science and Technologies Facilities Council of the United Kingdom; the Gordon 
and Betty Moore Foundation; the Heising-Simons Foundation; the French Alternative Energies and Atomic Energy Commission (CEA); 
the National Council of Science and Technology of Mexico; the Ministry of Economy of Spain, and by the DESI 
Member Institutions.  The authors are honored to be permitted to conduct astronomical research on Iolkam Du’ag (Kitt Peak), a mountain with 
particular significance to the Tohono O’odham Nation. 


\section*{Data availability}

The DESI Legacy Imaging Surveys used for this work is public at \url{https://www.legacysurvey.org/dr9/description/}. 




\bibliographystyle{mnras}
\bibliography{biblio} 






\bsp	
\label{lastpage}
\end{document}